\documentclass[a4paper,twocolumn,11pt,accepted=2023-10-19]{quantumarticle}

\pdfoutput=1
\usepackage[utf8]{inputenc}
\usepackage[english]{babel}
\usepackage[T1]{fontenc}
\usepackage{bm,bbold}
\usepackage{bbm}
\usepackage{amsfonts, amsmath, amsthm, amssymb} 
\usepackage{physics}
\usepackage{hyperref}
\usepackage{color}
\usepackage{subfig}
\usepackage{comment}
\usepackage[normalem]{ulem}
\usepackage[numbers,sort&compress]{natbib}

\begin{document}
\title{Certifying the quantum Fisher information from a given set of mean values: a semidefinite programming approach}

\author{Guillem M\"uller-Rigat}
\email{guillem.muller@icfo.eu}
\affiliation{ICFO - Institut de Ciencies Fotoniques, The Barcelona Institute of Science and Technology, 08860 Castelldefels (Barcelona), Spain}

\author{Anubhav Kumar Srivastava}
\affiliation{ICFO - Institut de Ciencies Fotoniques, The Barcelona Institute of Science and Technology, 08860 Castelldefels (Barcelona), Spain}

\author{Stanis{\l}aw Kurdzia{\l}ek}
\affiliation{Faculty of Physics, University of Warsaw, Pasteura 5, 02-093 Warszawa, Poland}

\author{Grzegorz Rajchel-Mieldzio{\'c}}
\affiliation{ICFO - Institut de Ciencies Fotoniques, The Barcelona Institute of Science and Technology, 08860 Castelldefels (Barcelona), Spain}

\author{Maciej Lewenstein}
\affiliation{ICFO - Institut de Ciencies Fotoniques, The Barcelona Institute of Science and Technology, 08860 Castelldefels (Barcelona), Spain}
\affiliation{ICREA, Pg. Llu\'{\i}s Companys 23, 08010 Barcelona, Spain}

\author{Ir\'en\'ee Fr\'erot}
\email{irenee.frerot@lkb.upmc.fr}
\affiliation{Univ Grenoble  Alpes, CNRS, Grenoble INP, Institut Néel, 38000 Grenoble, France}
\affiliation{Laboratoire Kastler Brossel, Sorbonne Universit\'e, CNRS, ENS-PSL Research University, Coll\`ege de France, 4 Place Jussieu, 75005 Paris, France}
\orcid{0000-0002-7703-8539}

\begin{abstract}
We introduce a semidefinite programming algorithm to find the minimal quantum Fisher information compatible with an arbitrary dataset of mean values. This certification task allows one to quantify the resource content of a quantum system for metrology applications without complete knowledge of the quantum state. We implement the algorithm to study quantum spin ensembles. We first focus on Dicke states, where our findings challenge and complement previous results in the literature. We then investigate states generated during the one-axis twisting dynamics, where in particular we find that the metrological power of the so-called multi-headed cat states can be certified using simple collective spin observables, such as fourth-order moments for small systems, and parity measurements for arbitrary system sizes.
\end{abstract}

\maketitle

\section{Introduction}
Quantum states of multipartite systems may represent useful resources for specific quantum technology applications \cite{deutsch2020}. It is therefore a central task for quantum information science to provide robust and reliable protocols to estimate the resource content of a quantum state $\hat \rho$ prepared in a given device. In several situations, the quantum advantage is a genuine quantum many-body effect, such as in quantum computing \cite{Preskill2018quantumcomputingin}, quantum simulation \cite{lewenstein2012ultracold,PRXQuantum.2.017003,fraxanet2022coming} or quantum metrology \cite{pezzeetal_RMP2018}. Hence, in the context of quantum certification -- namely, the task of estimating the resource content of a given experimental preparation --, a pervasive challenge is that of scalability. Given that quantum state tomography quickly becomes unfeasible for systems of increasing size, this challenge signifies that one must base the certification protocols on partial information, typically the knowledge of a few mean values of collective observables. A certification algorithm then represents a conservative approach, which quantifies the \textit{minimal} resource content over all quantum states compatible with the available partial information. A key property of all quantum resource quantifiers is that they are \textit{convex} functions $f(\hat \rho)$ of the quantum state; namely, a quantum resource \cite{RevModPhys.91.025001}, be it quantum coherence \cite{streltsovetal2017}, thermodynamic resources \cite{khanian2022resource}, quantum entanglement \cite{RevModPhys.81.865}, Einstein-Podolski-Rosen steering \cite{RevModPhys.92.015001}, Bell non-locality \cite{RevModPhys.86.419,scarani_book}, etc., cannot be produced (nor amplified) by mere statistical mixing of different preparations. Convexity turns out to be a key technical property for practical certification algorithms, as imposing compatibility of the quantum state with a given set of mean values --  namely, \textit{linear} constraints of the form ${\rm Tr}(\hat D_a \hat \rho)=\langle \hat{D}_a\rangle$ for $a=1,\dots,K$ where $\hat D_a$ are some observables forming the available data --, maintains the convexity of $f$. Therefore, the conservative estimate of the resource content of $\hat \rho$ compatible with the data can be cast altogether as a \textit{convex optimization} algorithm, hence presenting a unique and well-defined optimum. 

In this work, we focus on estimating the quantum Fisher information (QFI) associated to unitary transformations of $\hat \rho$ \cite{braunsteinCaves1994,bengtsson2007geometry}. The QFI quantifies the usefulness of the state for quantum metrology (namely, the task of measuring small parameters, such as magnetic fields or gravitational fields, beyond the capabilities of classical systems of the same size) \cite{pezzeetal_RMP2018}. While previous works have achieved significant progress in this respect \cite{apellanizetal2017,Gessner2019}, no existing approach fully exploits the convexity of the QFI to achieve an unbiased and certified estimate of the QFI given a set of mean values. In the present paper, exploiting the link of the QFI with the Uhlmann fidelity, we develop a semi-definite programming (SDP) algorithm to solve this problem. SDP represents a well-studied special class of convex optimization problems, for which efficient numerical routines are available \cite{bookSDP}. We apply our new approach to several problems of relevance to quantum metrology, such as Dicke states \cite{Zhang2014,pezzeetal_RMP2018}. We then explore the metrological resource produced during the out-of-equilibrium dynamics of the celebrated one-axis-twisting dynamics (OAT) \cite{Kitagawa1993}. In this case, we obtain the unexpected result that few-particle multi-headed cat states and their metrological power can be fully assessed by low-order moments of collective spin operators. We also find that Heisenberg-scaling metrological power can be optimally exploited by parity measurements throughout the OAT evolution. We finally illustrate our approach in small quantum spin chains with nearest-neighbour interactions.

\section{Semidefinite programming solution}
\noindent\textbf{Quantum Fisher information.--} In the context of quantum interferometry, the QFI evaluates the sensitivity of the state $\hat \rho$ to unitary transformations of the form $\hat\rho(\theta) = e^{-i\theta\hat G} \hat \rho e^{+i\theta\hat G}$ (linear encoding of the phase $\theta$) \cite{pezzeetal_RMP2018}. Here, $\hat G$ is the observable generating the unitary transformation. In the specific case of magnetometry, for instance, $\hat G$ is some component of the magnetic moment of the system, and $\theta$ would incorporate both the time of evolution and the strength of the magnetic field. Via the quantum Cramér-Rao bound \cite{Cramer1946,Rao1992,HELSTROM1967101}, the QFI ($F_Q[\hat \rho, \hat G]$) quantifies the ultimate sensitivity that can be achieved on the estimation of $\theta$ by making measurements on the state $\hat\rho(\theta)$, namely \cite{pezzeetal_RMP2018}: 
\begin{equation}
        (\Delta \theta)_{\rm est.}^2 \geq \frac{(\Delta \hat{O})^2}{(\partial_\theta \langle \hat O \rangle)^2}:=\Xi_{O}^2 \ge \frac{1}{F_Q[\hat{\rho},\hat{G}]}\ ~.
        \label{eq_Cramer_Rao}
    \end{equation}
Here, $(\Delta \theta)_{\rm est.}^2$ is the variance of the estimator of the phase $\theta$, and $\hat O$ is some observable whose mean value $\langle \hat O \rangle = {\rm Tr}[\hat O \hat \rho(\theta)]$ changes with $\theta$ as $\partial_\theta \langle\hat O\rangle = i\langle[\hat G,\hat O]\rangle$, and $(\Delta \hat O)^2=\langle \hat O^2 \rangle - \langle \hat O \rangle^2$ is the variance of $\hat O$. We denote by $\Xi_O$ the sensitivity of the observable $\hat O$ in the phase estimation task. The Cramér-Rao bound [Eq.~\eqref{eq_Cramer_Rao}] is valid for arbitrary observables $\hat O$. \\

\noindent\textbf{Statement of the certification problem.--} The problem we aim at solving is to certify the minimal QFI compatible with a set of $K$ expectation values $\langle\hat{\mathbf{D}} \rangle = (1,\langle\hat{D}_1 \rangle,\langle\hat{D}_2\rangle,...,\langle\hat{D}_K \rangle)$ (thereafter called data):
\begin{equation}
    \label{eq:central_problem}
    \begin{array}{crl}
    \min_{\hat{\rho} = \hat{\rho}^\dagger}& F_Q[\hat{\rho}, \hat G]&\\
    \mbox{s.t.}&  \mathrm{Tr}(\hat{\mathbf{ D}}\hat{\rho})& = \langle\hat{\mathbf{D}}\rangle \\
    &\hat{\rho} &\succeq 0
    \end{array} \\ ,
\end{equation}
where $\hat{\rho}$ is a quantum state acting in a Hilbert space $\mathcal{H}$ of a given dimension (namely: a Hermitian, semidefinite positive, unit-trace operator; note that we implement the unit-trace constraint with $\hat{D}_0 = \mathbb{I}$). 
As mentioned in the introduction, this problem is well-posed since the QFI is a convex function of the state and the linear constraints corresponding to the data maintain the convexity property.\\

\noindent\textbf{QFI and Uhlmann fidelity.--} The QFI can be related to a geometric notion of distance between the infinitesimally displaced states $\hat \rho_\pm = \hat \rho(\pm \delta \theta)$, namely \cite{afrnek2017, Zhou2019}: 
\begin{equation}
    \label{eq:fidelity_FQ}
    \mathcal{F}(\hat\rho_+,\hat\rho_- ) = 1-F_Q[\hat{\rho}, \hat{G}] (\delta \theta)^2 +  {\cal O}[(\delta\theta)^4]\ ,
\end{equation}
with $\sqrt{\mathcal{F}(\hat{\rho},\hat{\sigma})} = \mathrm{Tr}\left[\sqrt{\sqrt{\hat{\rho}}\hat{\sigma}\sqrt{\hat{\rho}}}\right]$ the Uhlmann fidelity. Importantly, if $\hat \rho$ is a pure state, then the QFI reduces to (four times) the variance of $\hat G$: $F_Q[|\psi\rangle\langle\psi|,\hat G] = 4[\langle\psi|\hat G^2|\psi \rangle -  \langle\psi|\hat G|\psi \rangle^2]:= 4(\Delta \hat{G})^2_{\ket{\psi}}$. 
As per Eq.~\eqref{eq:fidelity_FQ}, up to fourth-order corrections in $\delta\theta$, minimizing the QFI is equivalent to maximizing the fidelity between the states $\hat\rho(\pm\delta\theta)$ given the constraints $\mathrm{Tr}(\hat{\mathbf{D}}\hat{\rho}) = \langle \hat{\mathbf{D}}\rangle$.\\

\noindent\textbf{SDP solution to the certification problem.--} To arrive at our practical algorithm, we use the fact that the fidelity can be computed as a semidefinite-program (SDP) \cite{ Watrous, bookSDP},
\begin{equation}
    \begin{array}{crl}
    \sqrt{\mathcal{F}(\hat{\rho},\hat{\sigma})} = &\max_{\hat{L}}& \Re{\mathrm{Tr}(\hat{L})} \\
    &\mbox{s.t.}& \begin{pmatrix}
    \hat{\rho} & \hat{L}^\dagger \\
    \hat{L} & \hat{\sigma}
    \end{pmatrix} \succeq 0
    \end{array} \ .
\end{equation}
Furthermore, it is straightforward to impose compatibility of the state with the data in the SDP formulation, yielding the final SDP algorithm (for a given fixed and sufficiently small $\delta \theta$):
\begin{equation}
   \label{eq:sdp_FQ}
   \begin{array}{crl}
     \sqrt{{\cal F}_{\rm max}} =&\max_{\substack{\hat{L} \\ \hat{\rho} =\hat{\rho}^\dagger}}& \Re{\mathrm{Tr}(\hat{L})}  \\
    &\mbox{s.t.}& \begin{pmatrix}
    \hat{\rho}_+ & \hat{L}^\dagger \\
    \hat{L} & \hat{\rho}_-
    \end{pmatrix} \succeq 0 \\
    &&\hat{\rho}_{\pm} = \hat{\rho}\pm i\delta\theta[\hat{G}, \hat{\rho}] \\ 
     && \mathrm{Tr}(\hat{\mathbf{D}}\hat{\rho}) = \langle\hat{\mathbf{D}} \rangle  \\
     \end{array} \ .
\end{equation} 
The SDP algorithm delivers both the maximal fidelity and an optimal quantum state $\hat{\rho}^*$, whose QFI is evaluated either as $F_Q[\hat{\rho}^*,\hat{G}] = (1- {\cal F}_{\rm max})/\delta \theta^2 + {\cal O}(\delta\theta^2)$, or directly using the optimal state $\hat{\rho}^*$. In Appendix \ref{sec:app_dtheta}.1, we further discuss the accuracy of this certificate for small yet finite $\delta \theta$.\\

\noindent \textbf{Alternative SDP algorithm.--}
We propose an alternative SDP formulation to solve our central problem Eq.~\eqref{eq:central_problem}. The starting point is the definition of the inverse QFI as the minimal variance of a locally unbiased observable $\hat O$ \cite{helstrom1968minimum,zhou2023optimal}:   
\begin{equation}
\begin{array}{crl}
     \label{eq:QFI_var}
     F_Q [ \hat \rho, \hat G]^{-1}=&\min_{\hat{O} = \hat{O}^\dagger}& \mathrm{Tr}(\hat{\rho}\hat{O}^2)  \\
    &\mbox{s.t.}& \mathrm{Tr} \left( i [\hat \rho, \hat G] \hat O\right) = 1 \\
    &&\mathrm{Tr} \left( \hat \rho \hat O \right) = 0  \\
\end{array} \ .
\end{equation} 
Here, the operator $\hat O$ can be decomposed as $\hat O = \sum_i \tilde \theta(i) \ket{i} \bra{i}$, where $\left\{\ket{i}\right\}
$ is the measurement basis and $\tilde \theta(i)$ is the value of the estimator of $\theta$ associated with the $i$-th measurement outcome. The maximum of Eq.~\eqref{eq:QFI_var} over all states $\hat \rho$ compatible with a given set of expectation values $\langle\hat{\mathbf{D}} \rangle$ is equal to the inverse of the solution to problem Eq.~\eqref{eq:central_problem}. The minimization over $\hat O$, Eq.~\eqref{eq:QFI_var}, can be reformulated as a maximization problem using the strong duality theorem for SDP.  Finally, the resulting double maximization problem (over $\hat \rho$ and over variables dual to $\hat O$) can be recasted as a single SDP---see Appendix \ref{sec:app_dtheta} for the derivation and exact formulation of this SDP, and for the comparison with the fidelity based approach. \\

\noindent\textbf{Exploiting the metrological usefulness.--} After obtaning the minimal QFI for fixed arbitrary data $\hat{\mathbf{D}}$, one may ask how to operationally exploit the corresponding metrological resource. In other words, one aims at finding an explicit observable $\hat O$ with sensitivity $\Xi_{O}^{-2}$ as close as possible to the QFI. From the optimal state $\hat{\rho}^*$, such observable can be provided with the certificate. This privileged observable corresponds to the symmetric logarithmic derivative (SLD) \cite{pezze2014quantum} $S$ which is derived from the relation $i[\hat{G},\hat{\rho}^*] = \{\hat{S},\hat{\rho}^* \}/2$. For pure states, $\hat{\rho}^* = \ketbra{\Psi}$, $\hat{S} = 2i[\hat{G},\ketbra{\Psi}]$. For this observable, $\Xi_{S}^{-2} = \partial_\theta \langle S \rangle = \langle i [\hat{G}, \hat{S}] \rangle =\langle \hat{S}^2\rangle =F_Q[\hat{\rho}^*,\hat{G}]$ is verified. Notice, however, that in practice the SLD observable might be impractical to measure, and the practical optimal observable must be found on a case-by-case basis \cite{Gessner2019}.\\

\noindent\textbf{Taking advantage of symmetries.--} While being a convex-optimization algorithm, the SDP of Eq.~\eqref{eq:sdp_FQ} involves the complete density matrix $\hat\rho$ as a variable of the problem, and hence is hardly scalable beyond about $N=10$ qubits. However, in several cases of practical interest, both the generator $\hat G$ and the observables $\hat{\mathbf{D}}$ respect elementary symmetries, such as translation invariance, or even invariance under permutation of the elementary degrees of freedom. Such symmetries can be accommodated in the SDP by splitting the optimization into different symmetry sectors, which yields a more scalable algorithm. As further discussed in Appendix \ref{sec:app_symmetries}, if both the observables $\hat{\mathbf{D}}$ and the generator $\hat G$ obey a certain symmetry, it is sufficient to optimize over symmetric states of the form $\hat{\rho} = \oplus_{J}\hat{\varrho}^{(J)}$, where $\hat\varrho^{(J)}$ is a (non-normalized) quantum state acting on the symmetry sector $J$ (with ${\rm Tr}[\hat\varrho^{(J)}]$ the probability to find the system in the symmetry sector $J$). In essence, taking advantage of the symmetries allows one to block-diagonalize the problem with respect to the symmetry sectors, whose dimensions are typically much smaller than the complete many-body Hilbert space, yielding an SDP algorithm with a more favorable scaling (see Appendix \ref{sec:app_symmetries} for a specific example on permutation invariance). \\

\noindent\textbf{Alternative approaches.--} The central problem Eq.~\eqref{eq:central_problem} has appeared in different contexts throughout the literature. In Appendix \ref{sec:app_apellaniz}, we give a concise review of past efforts to solve it and discuss its comparison to the algorithms proposed here. \\

\noindent\textbf{Beyond unitary evolutions.--} In this paper, we focus on situations where the metrological protocol consists in a unitary evolution of the probe: $\hat\rho(\theta) = e^{-i\theta\hat G} \hat \rho e^{+i\theta\hat G}$. However, the SDP of Eq.~\eqref{eq:sdp_FQ} can be extended to a generic (namely: non-unitary) parameter encoding operation by defining $\hat{\rho}_{\pm}(\theta) = \hat{\rho}(\theta)\pm \delta\theta\hat{\mathcal{L}}[\hat{\rho}(\theta)]$, where $\hat{\mathcal{L}}$ is the corresponding Lindbladian, $\partial_\theta\hat{\rho}(\theta) = \hat{\mathcal{L}}[\hat{\rho}(\theta)]$. This can be used to incorporate noise and other imperfections of the metrological task \cite{Len2022}, or to model more general open-system dynamics. Notice that in this case, the QFI becomes $\theta$-dependent.\\

In the remainder of the paper, we illustrate our approach by providing the certified metrological usefulness in several classes of states of experimental relevance. We focus on states of an ensemble of $N$ quantum spin-1/2 particles. In Section \ref{sec:Dicke_states}, we consider Dicke states, which in contrast to spin-squeezed states have a vanishing mean spin. In Section \ref{sec:OAT}, we then consider the family of states generated during the so-called one-axis twisting dynamics \cite{Kitagawa1993} using collective spin observables of increasing orders.

\section{Metrological usefulness of unpolarized states}
\label{sec:Dicke_states}
In this section, we focus on unpolarized states, namely states of vanishing mean spin: $\langle \hat {\bf J} \rangle = 0$ with $\hat {\bf J} = \sum_{i=1}^N \hat {\bf S}^{(i)}$ and $\hat {\bf S}^{(i)}=(S_x^{(i)}, S_y^{(i)}, S_z^{(i)})$ individual spin-1/2 operators. Such unpolarized states may offer a metrological advantage, the most prominent example being balanced Dicke states, such that $\langle \hat{\bf J}^2\rangle = (N/2)(N/2+1)$ (maximal total spin) and $\langle \hat J_x^2 \rangle = 0$ (eigenstate of the collective spin along $x$ with eigenvalue $0$) \cite{HollandPRL1993}. The balanced Dicke state displays an exquisite sensitivity to rotations around any axis in the $yz$ plane, with QFI given by $F_Q[\hat \rho, \hat J_z]=4\langle \hat J_z^2\rangle = N(N/2+1)$. To exploit this sensitivity, a possible approach is to measure $\langle\hat J_x^2\rangle(\theta)$, allowing one to reach the sensitivity bound of the QFI \cite{KimPRA1998,luckescience2011,Apellaniz2015,zouetal2018}. In this protocol, one needs to estimate the variance of $\hat J_x^2$, and hence to access fourth-order moments of the collective spin. Here, we are instead interested in quantifying the metrological usefulness of unpolarized states using only second moments of collective spins. Our motivation for doing so is that in an experiment, it is impossible to prepare the balanced Dicke state with unit fidelity, and it is natural to certify the QFI using only the knowledge of $\langle \hat{J}_x^2\rangle$, $\langle \hat{J}_y^2\rangle$, $\langle \hat{J}_z^2\rangle$ . \\

\noindent\textbf{Case $\langle \hat J_x^2 \rangle=0$.--} In particular, for $\langle\hat{J}_x^2\rangle = 0$, as we prove in Appendix \ref{sec:app_obs1}, the QFI with respect to the generator $\hat{J}_{{\bf n}_\perp}$ is equal to $4\langle \hat J_{{\bf n}_\perp}^2 \rangle$, where ${\bf n}_{\perp}$ is any direction in the $yz$ plane. The constraint $\langle \hat{J}_x^2\rangle = 0$ implies that any state compatible with it is an eigenstate of $\hat{J}_x$ and therefore it is rotationally invariant around the $x$ axis. Consequently, $\langle\hat{J}_y^2 \rangle = \langle\hat{J}_z^2 \rangle = \langle \hat{\mathbf{J}}^2\rangle/2 $. Hence, such observation extends the property of the balanced Dicke state (which is a pure state of maximal total spin $\langle \hat{\mathbf{J}}^2\rangle=(N/2)(N/2+1)$) to general mixed states with arbitrary total spin. \\

\noindent\textbf{Case $\langle\hat J_x^2 \rangle>0$.--} We then investigate the effect of a finite but small nonzero value of $\langle \hat{J}_x^2 \rangle$ on the QFI of $\hat J_z$.  In Appendix \ref{sec:app_Dicke_scaling}, we study the minimal $F_Q(\hat J_z)$ compatible with a given $\langle \hat J_x^2 \rangle$ and a given total spin $\langle {\bf J}^2 \rangle \sim N^2$. Our main conclusion is that for $\langle\hat{J}_x^2 \rangle\gtrsim 1/4$, the Heisenberg scaling $F_Q(\hat J_z) \sim N^2$ is lost. This result suggests that certifying the metrological power of Dicke-like states requires knowledge beyond second moments. \\ 

\begin{figure}[h]
\hspace*{-0.5cm}
  \includegraphics[width=0.5\textwidth]{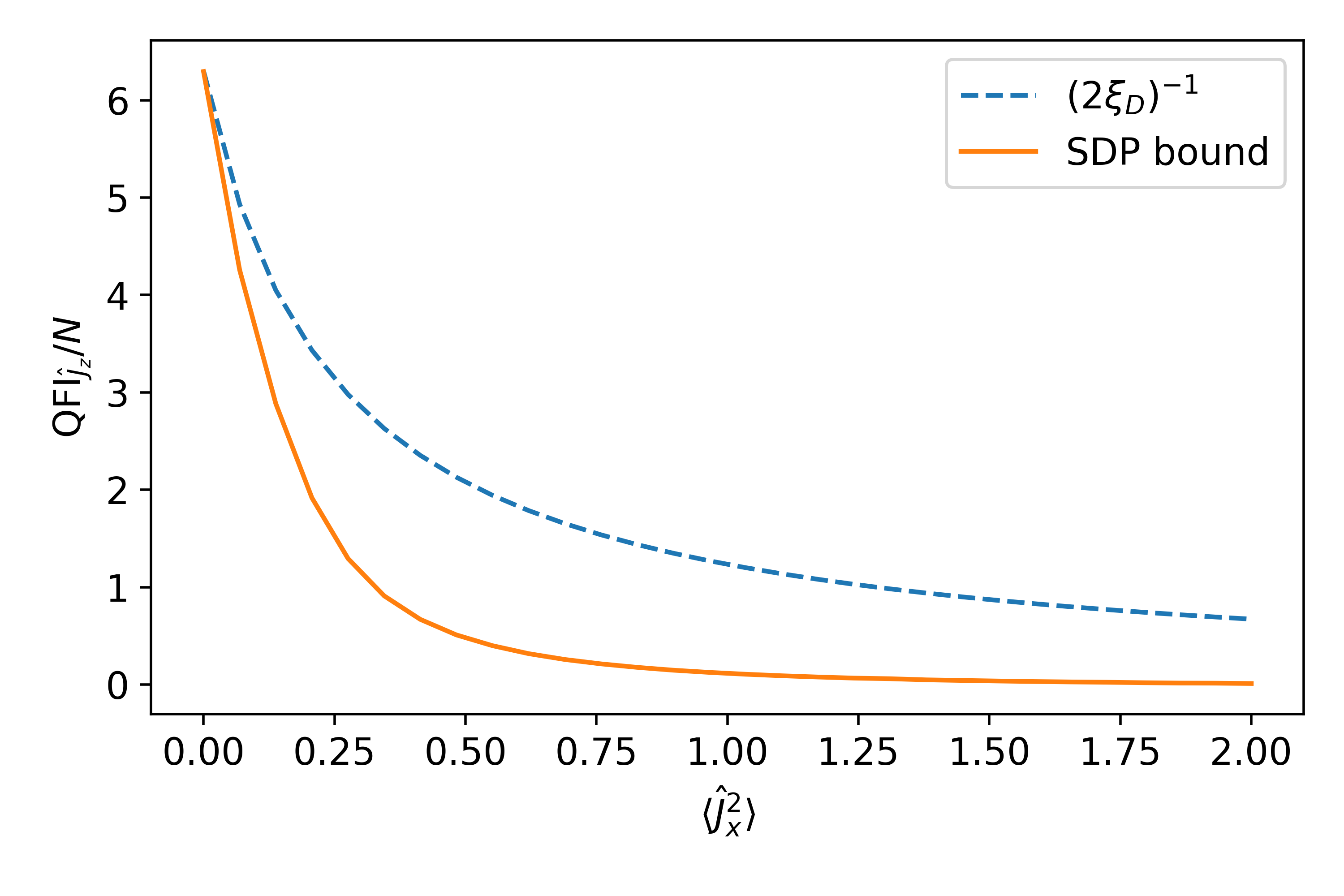} 
	\caption{For each value of $\langle\hat{J}_x^2 \rangle$ we obtain the minimum QFI compatible with $\langle\hat{J}_x \rangle=0, \langle\hat{J}_x^2 \rangle, \langle\hat{J}_y^2+\hat{J}_z^2 \rangle = 0.7(N/2)(N/2+1)- \langle\hat{J}_x^2 \rangle$, and $N=16$ (solid orange line). We compare it with the Dicke squeezing parameter [Eq.~\eqref{eq:def_SSD}] (dashed blue line).}
  \label{fig:Dicke_param}
\end{figure}

\noindent\textbf{Questioning the Dicke squeezing parameter.--} Here we question the validity of a proposed parameter to estimate the metrological usefulness of unpolarized states, namely the so-called Dicke squeezing parameter introduced in Ref.~\cite{Zhang2014}:
\begin{equation}
   \label{eq:def_SSD}
    (2\xi_D)^{-1} = \frac{\langle \hat{J}_y^2 + \hat{J}_z^2\rangle}{N(2\langle \hat{J}_x^2\rangle+1/2)} ~.
\end{equation}
This parameter was observed numerically to quantify the metrological usefulness of a specific class of states (namely: superpositions of Dicke states close to the balanced Dicke state), but to our knowledge, no proof of the general relevance of the Dicke squeezing parameter has been given in the literature. Note that, as per the previous observation, for $\langle \hat{J}_x^2\rangle = 0$ one has that $\mathrm{QFI}_{\hat{J}_z}/N = (2\xi_D)^{-1}$. Our approach allows us to provide a rigorous lower bound to the QFI, as opposed to $\xi_D$ which only provides a conjectured lower bound based on numerical observations. In Figure \ref{fig:Dicke_param}, we compare $(2\xi_D)^{-1}$ with the minimum QFI of $\hat{J}_z$ compatible with a reduced total spin of $\langle\hat{\mathbf{J}}^2 \rangle=0.7(N/2)(N/2+1)$ as a function of $\langle \hat{J}_x^2\rangle$.

We find that the Dicke squeezing parameter [Eq.~\eqref{eq:def_SSD}] only captures the metrological usefulness for very small values of $\langle \hat J_x^2 \rangle$, but in general it largely overestimates the QFI of the state. Therefore, the main outcome of this study is that this parameter cannot be used as a faithful estimate of the metrological usefulness of the system without further assumptions. Notice though that in Ref. \cite{Apellaniz2015}, a proper squeezing parameter tailored to Dicke states has been derived, which constitutes a valid lower bound of the QFI, and allows one to strongly improve the estimation of the metrological usefulness of the state. However, this criterion involves fourth moments of the collective spin, while our goal here was to include data up to the second moments only. It would be interesting in a future work to investigate the tightness of the squeezing parameter of Ref.~\cite{Apellaniz2015} by including fourth-order moments in our SDP algorithm.

\section{Metrological usefulness of one-axis twisting-generated states}
\label{sec:OAT}

In this section we further apply our certification algorithm by considering data produced during the one-axis twisting (OAT) dynamics starting from a state polarized along the $x$ axis \cite{Kitagawa1993}:
\begin{equation}
  \label{eq:OAT}
    \ket{\Psi(t)} = e^{it\hat{J}_z^2}\ket{\theta = \pi/2,\phi = 0} \ ,
\end{equation}
where $\ket{\theta,\phi} = [\cos(\theta/2)\ket{\uparrow} + e^{i\phi} \sin(\theta/2)\ket{\downarrow}]^{\otimes N}$ is a coherent spin state (CSS) of length $J = N/2$, with $(\theta,\phi)$ the usual polar and azimuthal angles. 

The OAT protocol has been widely investigated both theoretically and experimentally to prepare metrologically useful entangled many-body states in atomic ensembles \cite{pezzeetal_RMP2018}. Its realization in different regimes of structured systems such as optical lattices has also been explored in the context of cold-atom and Rydberg-atom quantum simulators  \cite{Podzie2020,HernndezYanes2022,Podzie2022,PhysRevLett.129.150503,Dziurawiec2023,bornet2023scalable}. The OAT dynamics generates spin squeezing at short times $t\lesssim 1/\sqrt{N}$ , with strength $\xi^{-2}_{R} \sim N^{3/2}$. At longer times, the mean spin $\langle \hat{J}_x\rangle$ vanishes and quantum correlations are not simply reflected in spin squeezing. While the theoretical metrological usefulness continues to increase with time, it becomes more demanding to certify or exploit it, as the correlations are spread in higher moments of the collective spin fluctuations. In particular, at times $t = \pi/q$, the state is a so-called $q$-headed cat state (MHCS), namely an equal-weight superposition of $q$ CSS in the $xy$ plane ($\theta = \pi/2$) at angles $\phi_m = 2\pi m/q$ for $m \in \{0,1,..,q-1 \}$ \cite{Agarwal1997, Song2019}. This includes, for $t=\pi/2$, the conventional cat state (with two heads) marking half of the total period of the dynamics. Before discussing the metrological usefulness of MHCS, and more general states generated during the OAT dynamics, we briefly review the case of spin-squeezed states.\\

 \noindent\textbf{Spin-squeezed states.--} Spin-squeezed states represent the most widely studied class of many-body spin states useful for metrology, and are generated at short time in the OAT dynamics. Those states have a nonzero mean spin $\langle \hat {\bf J} \rangle = \langle \hat J_{\bf n} \rangle \neq 0$ (with ${\bf n}$ the direction of the mean spin) and a reduced variance for a spin component transverse to the mean spin $\langle \hat J_{{\bf n}_\perp}^2 \rangle < \langle \hat J_{\bf n} \rangle^2 / N$ (with ${\bf n}_\perp$ orthogonal to ${\bf n}$) \cite{winelandetal1994}. For simplicity, let us assume that the mean spin is along $x$ (as is the case in the OAT dynamics), and that the minimal variance orthogonal to $x$ is along $s$ (squeezing axis). The QFI for rotations around $as = s\times x$ (antisqueezed axis) is lower-bounded by the spin squeezing parameter \cite{pezzeetal_RMP2018}:
\begin{equation}
\label{eq:spin_squeezing_y}
    \xi_R^{-2} = \frac{\Xi_{J_s}^{-2}}{N} = \frac{\langle \hat J_x \rangle^2}{N\langle \hat J_s^2 \rangle} \le \frac{F_Q[\hat \rho, \hat J_{as}]}{N} ~.
\end{equation}
It is natural to investigate the tightness of this bound, namely, to find the minimal $F_Q[\hat\rho, \hat J_{as}]$ compatible with given $\langle \hat J_x \rangle$ and $\langle \hat J_s^2\rangle$. This question has been investigated previously in Ref.~\cite{apellanizetal2017} (see Appendix \ref{sec:app_apellaniz} for a discussion of the algorithm used in this paper), and the bound is known to be tight except for data points at the boundary of the feasible set (namely, the set of values for $\langle \hat J_x \rangle$, $\langle \hat J_s^2\rangle$ for which there exists a quantum state compatible with them), where the bound is tight within $\sim 2\%$ accuracy.\\

In the remainder of this section, we will use expectation values as produced during the OAT dynamics [Eq.~\eqref{eq:OAT}] to derive, using our SDP algorithm, the sensitivity bound to collective spin rotations. We will compare our obtained bounds with well-known lower bounds of the QFI, such as spin squeezing [Eq.~\eqref{eq:spin_squeezing_y}], or non-linear generalizations thereof \cite{Gessner2019}.\\

\begin{figure}[h]
\hspace*{-0.5cm}
   \includegraphics[width=0.5\textwidth]{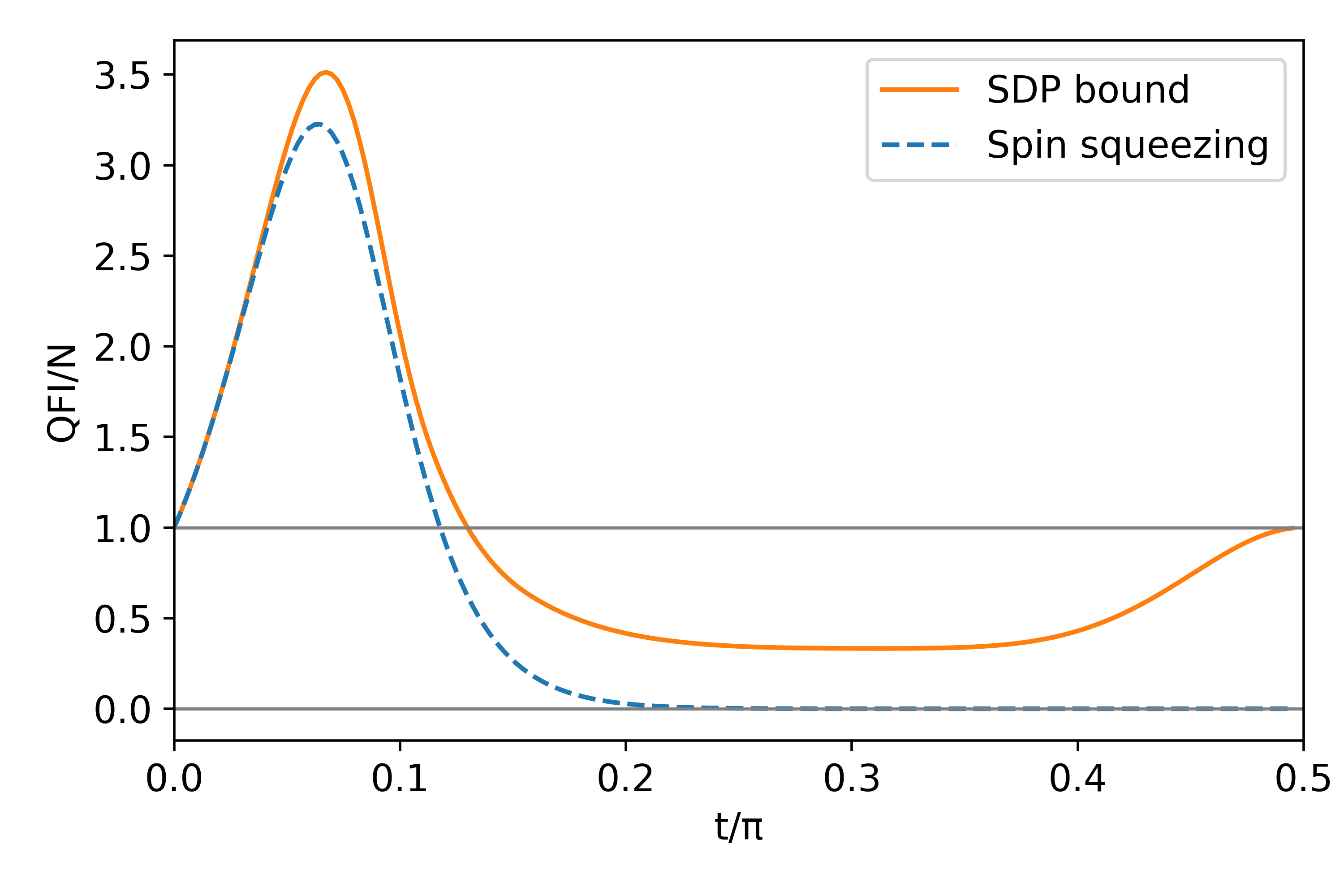}
   \caption{Evolution of the metrological bounds from second moments of the collective spin during OAT dynamics for $N=10$ particles. From bottom to top: in blue (dashed) conventional spin squeezing parameter and in orange (solid line), the SDP bound based on the same data. We also depict horizontal lines at $\mathrm{QFI} = 0$ and  $\mathrm{QFI} = N$ to represent no metrological usefulness and the coherent state limit respectively.       }
  \label{fig:OAT_lin}
\end{figure}

\noindent\textbf{Second moments.--} We start by taking the first and second moments of collective spin observables as input data to our algorithm, namely $\langle\hat{J}_a \rangle$, and $\mathrm{Re}\{\langle\hat{J}_a \hat{J}_b\rangle\}$ for $a,b\in\{x,y,z\}$. Note that since $\langle \hat {\bf J}^2 \rangle=(N/2)(N/2+1)$ is given explicitly by the data, it is sufficient to optimize over the totally symmetric sector whose dimension is $N+1$, instead of the complete Hilbert space of dimension $2^N$. Such expectation values contain also implicitly the spin squeezing bound of Eq.~\eqref{eq:spin_squeezing_y}, where the squeezed component $\hat{J}_s$ is used to sense rotations generated by $\hat G = \hat{J}_{as}$. At short times, the orientations $\{s,as\}$ are contained in the $yz$ plane (orthogonal to the mean spin, which is along $x$) and depend on time. At longer times, the state wraps around the Bloch sphere and becomes unpolarized. In this case, we use as a generator $\hat G = \hat J_y$. Minimizing the QFI with respect to the generator $\hat{G}$ allows us to compare both our SDP bound and the spin-squeezing bound. The result is displayed in Figure \ref{fig:OAT_lin}.

The results presented in Figure \ref{fig:OAT_lin} lead us to several important observations. First, there is a gap between our bound and the spin squeezing parameter, demonstrating that using the same second moments as used to build the optimal spin-squeezing parameter, one can in fact certify more metrological usefulness than predicted by the same squeezing parameter. As a matter of fact, adding to the data the second moment along the mean spin ($\langle \hat{J}_x^2\rangle$) in addition to the mean spin itself ($\langle\hat{J}_x \rangle$) and to the squeezed second moment ($\langle\hat{J}_{s}^2 \rangle$) is already sufficient to open this gap between the spin squeezing parameter and the SDP bound, and gives the same QFI lower-bound as using the full correlation matrix of the collective spin. We note that an illustrative special case is at $t =\pi/2$. At this time, a cat (GHZ) state is created $\ket{\mathrm{GHZ}} = (\ket{\pi/2, 0} + e^{i\Theta}\ket{\pi/2, \pi})/\sqrt{2}$. As it has a vanishing mean spin, its spin squeezing bound is zero. It features $\langle\hat{J}_x^2 \rangle = N^2/4,\langle\hat{J}_y^2 \rangle = \langle\hat{J}_z^2 \rangle = N/4$, which implies that it exhibits maximal QFI with respect to rotations generated by $\hat{J}_x$, $F_Q[\ket{\mathrm{GHZ}}, \hat{J}_x] = 4(\Delta \hat{J}_x)^2 = N^2$. However, using only second moments it is not possible to certify this metrological usefulness, as they are always compatible with a state invariant under rotations around $x$ such as a statistical mixture of two coherent spin states $[\mathrm{GHZ}]=(\ketbra{\pi/2,0} + \ketbra{\pi/2,\pi})/2$. Nonetheless, one can certify the metrological usefulness of rotations around $y$ (or $z$), for which $F_Q[[\mathrm{GHZ}], \hat{J}_y] = N$, which is equivalent to considering rotations of a coherent spin state polarized along $\pm x$. Finally, we notice that more generally, at all times at which the mean spin is zero, our SDP algorithm certifies metrological usefulness for rotations around $y$ with a QFI lower bound $\sim N $ (see Appendix \ref{sec:app_OAT_scaling} for details).    

In order to further reveal the metrological usefulness of non-spin-squeezed states beyond the small QFI lower-bound reported in Figure \ref{fig:OAT_lin}, it is necessary to consider data involving higher moments of the collective spin.\\

\begin{figure}[h]
   \hspace*{-0.5cm}
   \includegraphics[width=0.5\textwidth]{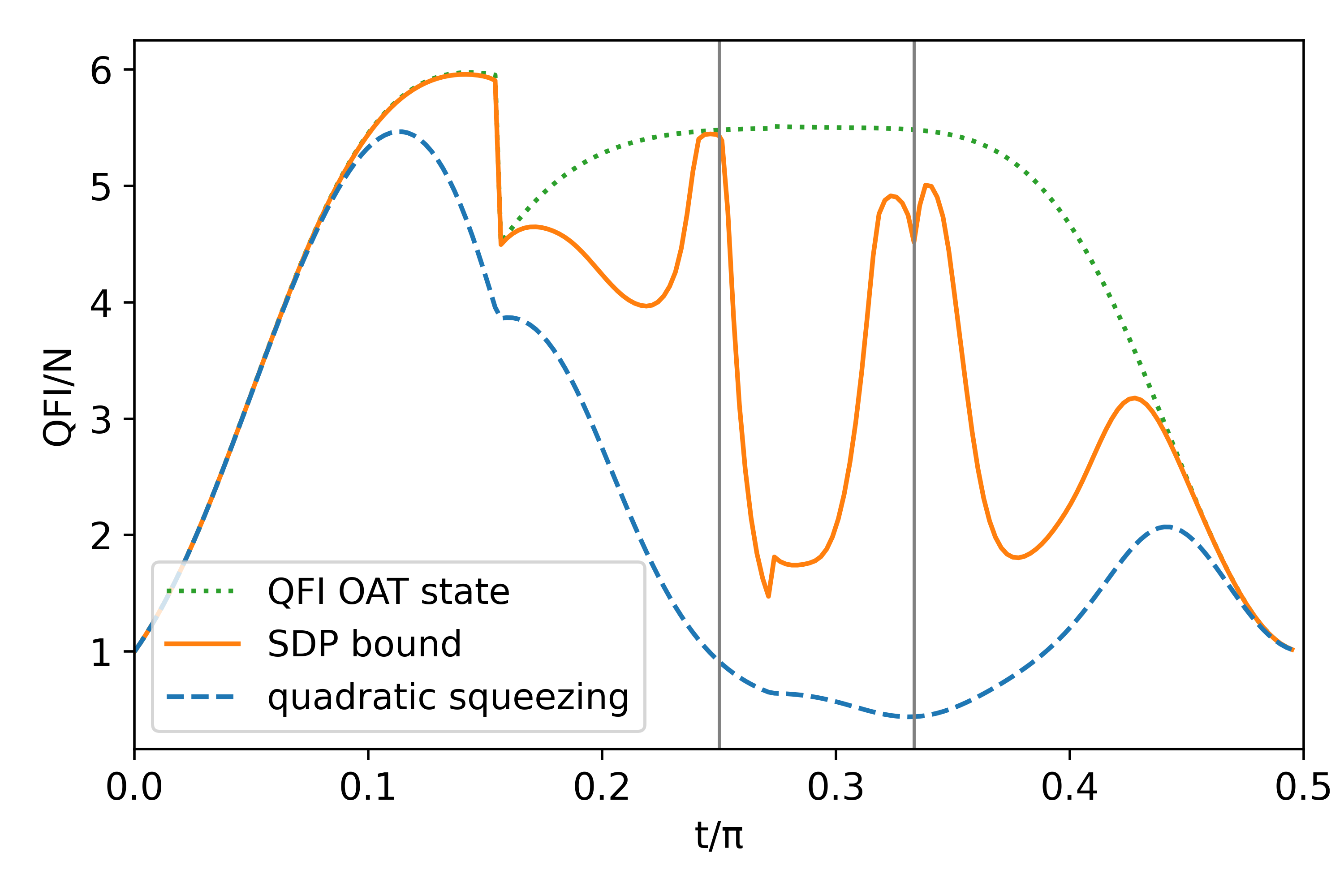} 
   \caption{Evolution of the metrological bounds from fourth moments of the collective spin during OAT dynamics for $N=10$ particles. From bottom to top: in blue (dashed) nonlinear spin squeezing parameter as introduced in Ref.~\cite{Gessner2019}, in orange (solid line), the SDP bound, based on the same data. The green dotted line is $4(\langle \hat G^2 \rangle - \langle \hat G \rangle^2)$ and corresponds to the QFI of the OAT-evolved state $\ket{\Psi(t)}$ of Eq.~\eqref{eq:OAT} with respect to the same generator $\hat{G}$ as used to compute the other bounds.}
  \label{fig:OAT_quad}
\end{figure}

\noindent\textbf{Fourth moments.--} We then add to our SDP algorithm the information on third and fourth moments: $\mathrm{Re}\{\langle \hat{J}_a\hat{J}_b\hat{J}_c \rangle\}$, $\mathrm{Re}\{\langle \hat{J}_a\hat{J}_b\hat{J}_c\hat{J}_d\rangle\}$ for $a,b,c,d\in\{x,y,z\}$, in addition to the data considered in Figure \ref{fig:OAT_lin}. We emphasize that in Ref. \cite{Gessner2019}, a method is proposed to incorporate systematically such quantities in generalized squeezing bounds, quantifying the optimal sensitivity to collective spin rotations for all observables in the span of the data operators. Namely, if the data include mean values of the form $\{\langle\hat O_a \rangle\}_a = \langle\hat{\mathbf{O}}\rangle$ and second moments of it $\{\langle\hat{O}_a\hat{O}_b \rangle\}_{a,b} = \langle\hat{\mathbf{O}}\hat{\mathbf{O}}^T \rangle \}$, then Ref.~\cite{Gessner2019} allows one to build the optimal observable of the form $\hat O = \mathbf{v}\cdot\hat{\mathbf{O}}$ such that $\Xi_O^{-2}$ is maximal, as well as the optimal direction for collective spin rotations. Here, if one considers $\hat{\mathbf{O}}$ up to second moments, using fourth-order moments in $\hat{\mathbf{D}}$ is sufficient to derive the corresponding bound. The details of the approach of Ref.~\cite{Gessner2019} can be found in Appendix \ref{sec:app_apellaniz}. Using the said approach, we find that at short times, the optimal generator is $\hat{G} = \hat{J}_{as}$ (as was already the case using only up to second moments of the collective spin); then until $t\lesssim 0.15\pi$ the optimal generator is $\hat{G} = \hat{J}_x$; then for $t \gtrsim  0.27\pi$ the optimal generator is $\hat{G} = \hat{J}_y$. We compare the non-linear spin squeezing parameter obtained using the method of Ref.~\cite{Gessner2019} to the QFI lower-bound obtained by our SDP, using up to fourth-order moments, and for the same optimal generator. Notice that in general the optimal generator as given by Ref.~\cite{Gessner2019} need not be the same as the one for our approach. The result of the comparison of both sensitivity bounds for $N=10$ particles is shown in Figure \ref{fig:OAT_quad}. We also plot the QFI of the OAT state $\ket{\Psi(t)}$ with respect to $\hat G$.  (Note that in Figure~\ref{fig:OAT_quad}, we have added infinitesimal noise to the quantum state which modifies the data as $\langle\hat{\mathbf{D}}\rangle\mapsto (1-p)\langle\hat{\mathbf{D}}\rangle + p\mathrm{Tr}(\hat{\mathbf{D}})/(N+1)$ with $p =10^{-4} $; this has a negligible effect on the QFI of the state or the generalized squeezing bound, and avoids singular results in the SDP bound.)

We observe consistently that our bound systematically improves over the non-linear squeezing parameter of Ref.~\cite{Gessner2019} as soon as the non-gaussianity of the state becomes significant (that is, at times $t/\pi \gtrsim 0.1$). However, we notice that in contrast to the previous case considering only up to second-order observables, here the gap appears to close when increasing $N$ (see Appendix \ref{sec:app_OAT_scaling} for a detailed discussion on the scaling). Furthermore, a striking fact is that close to times $t = \pi/4, \pi/3$, where MHCS are realized, our bound nearly saturates the full QFI of the state. This is remarkable, for MHCS are highly non-gaussian states and are rather characterized by high-order interference fringes stemming from the coherent superpositions of several well-separated coherent spin states, while our bound requires using spin fluctuations only up to order four. We observe that for the 4-MHCS at time $t=\pi/4$ and for $N=2,4,6,8,10$, our bound actually reaches the maximal QFI, namely $N(N+1)/2$. This result stems from the unexpected fact that by fixing the collective spin fluctuations up to order four, one actually fully specifies the state: there exists an operator $\hat W$, built as a linear combination of the data, such that the state corresponds to its extremal eigenvalue: $\min_{\hat{\rho}} {\rm Tr}(\hat \rho \hat{W}) = {\bra{4\mathrm{MHCS}}} \hat{W} {\ket{4\mathrm{MHCS}}} = \lambda_{\min}(\hat{W})$, where $\lambda_{\min}$ denotes the minimum eigenvalue of $\hat W$ with corresponding eigenvector $\ket{4\mathrm{MHCS}}$. In Appendix \ref{sec:app_witness} we outline a systematic procedure to find such operator (if it exists). However, we observe that this result is a finite-size effect, only valid in few-particle systems: indeed, as the number of spins is increased, the peaks for our lower-bound to QFI at times $t=\pi/4, \pi/3$ drop. For the 4-MHCS, the certified QFI drops below $N$ (coherent-state limit) for $N=24$. Regarding the 3-MHCS, the certification drops below $N$ already for $N = 10$ (see Appendix \ref{sec:app_OAT_scaling} for further details). For larger particle numbers, higher order moments then become necessary to certify the metrological resource of highly non-gaussian states such as MHCS. \\

\begin{figure}[h]
   \hspace*{-0.5cm}\includegraphics[width=0.5\textwidth]{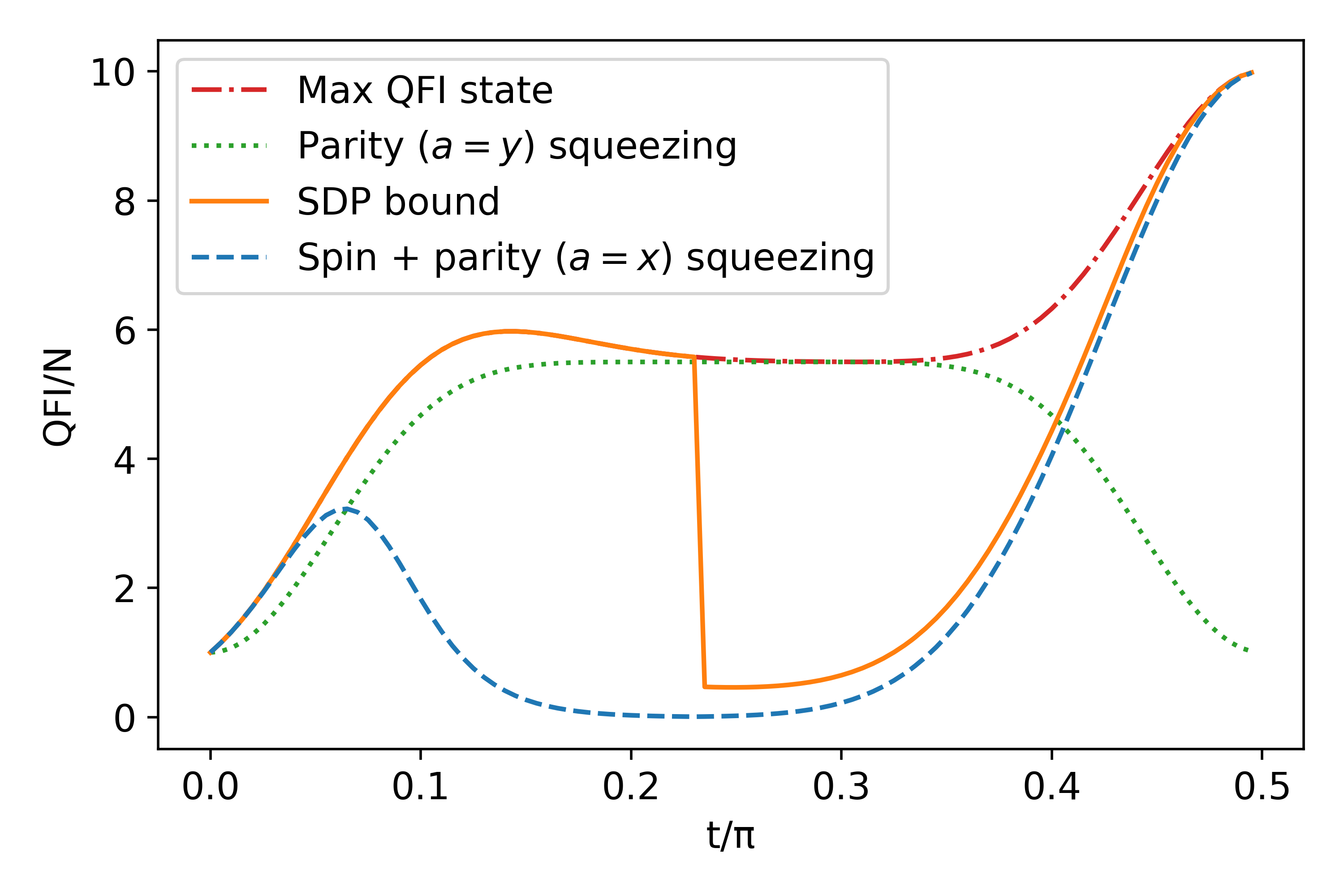} 
   \caption{Evolution of the metrological bounds from second moments of spin and parity measurements during OAT dynamics for $N=10$ particles. From bottom to top: in blue (dashed) non-linear squeezing parameter derived from Ref.~\cite{Gessner2019} using $\hat{\mathbf{O}} = \{\hat{J}_x, \hat{J}_y, \hat{J}_z, \hat{\Pi}_x, \hat{\Pi}_y, \hat{\Pi}_z \}$ as measurement operators; orange curve (solid line) -- the corresponding SDP bound using the same data. In green (dotted) parity squeezing parameter of Eq.~\eqref{eq:parity_squeezing} with $a=y$ (generator direction) and $b=x$ (measurement direction). In red (dashed-dotted) is plotted the maximal QFI of the OAT-generated states with respect to collective spin rotations.}
  \label{fig:OAT_linpar}
\end{figure}

\noindent\textbf{Parity measurements.--} As a last illustration of our algorithm, we consider the use of parity operators $\hat{\Pi}_a = (-1)^{\hat{J}_a + N/2}$, which are paradigmatic operators to exploit the maximal metrological usefulness of the cat (GHZ) state generated at time $t=\pi/2$ in the OAT dynamics \cite{Leibfried2005}. In analogy with Eq.~\eqref{eq:spin_squeezing_y}, one can construct the parity squeezing parameter: 
\begin{equation}
\label{eq:parity_squeezing}
\Xi^{-2}_{\Pi} = \frac{|\langle [\hat{J}_{a}, \hat{\Pi}_{b}]\rangle|^2}{1- \langle \hat{\Pi}_b\rangle^2} \ .
\end{equation}
For the GHZ state, $\ket{\mathrm{GHZ}} = (\ket{\pi/2, 0}+e^{i\Theta}\ket{\pi/2,\pi})/\sqrt{2}$, the optimal axes are $a=x$ and $b$ contained in the $yz$ plane, yielding $\Xi^{-2}_{\Pi}=N^2$, which is the maximal sensitivity allowed by quantum mechanics (the so-called Heisenberg limit). Such a result can be also obtained with the approach of Ref.~\cite{Gessner2019} using as operators $\hat{\mathbf{O}} = \{\hat{J}_x, \hat{J}_y, \hat{J}_z, \hat{\Pi}_x, \hat{\Pi}_y, \hat{\Pi}_z \}$, and including their expectation values and the correlation matrix $\langle\hat{\mathbf{O}}\hat{\mathbf{O}}^T \rangle$. Using this approach, the optimal generator throughout the OAT dynamics is in the anti-squeezed direction $\hat{J}_{as}$ for times $t\lesssim 0.27\pi$, and then $\hat{J}_x$ for longer times. As in the previous illustrations, our proposed SDP-based algorithm is used to find a stronger metrological sensitivity based on the same data and with respect to such optimal generator. In Figure \ref{fig:OAT_linpar} the corresponding bounds are compared. There, we depict the maximal QFI associated with collective spin rotations, that is, (four times) the largest eigenvalue of the covariance matrix $C_{ab} = \mathrm{Re}\{\langle \hat{J}_a\hat{J_b} \rangle\} - \langle \hat{J}_a\rangle\langle\hat{J}_b \rangle$ for $a,b\in\{x,y,z\}$. 

We observe that at the times where the generator is $\hat{J}_{as}$, our bound captures the full QFI of the state, a strong improvement with respect to the prediction of Ref.~\cite{Gessner2019}. In particular, we verify that it is sufficient in our algorithm to input the data $\langle\hat{\Pi}_x \rangle, \mathrm{Im}\{\langle\hat{\Pi}_x\hat{J}_{as} \rangle\}$ to saturate the full QFI of the state. In addition, we have verified that this QFI can be saturated using simply the parity squeezing parameter introduced in Eq.~\eqref{eq:parity_squeezing}, with directions $a= as, b= x$. This simple observation highlights the possibility to use parity operators to exploit the maximal sensitivity during the whole OAT dynamics before the generation of the GHZ state.  

At longer times ($t\gtrsim 0.27\pi$), when the optimal generator as predicted by Ref.~\cite{Gessner2019} is $\hat J_x$, our bound only yields a modest improvement with respect to the non-linear squeezing parameter of Ref.~\cite{Gessner2019}. However, we notice our bound can be drastically improved up to times $t\lesssim 0.35\pi$, again saturating the full QFI of the state, by just changing the generator from $\hat J_x$ to $\hat J_y$. Also in this case, the bound is saturated by the parity squeezing parameter of Eq.~\eqref{eq:parity_squeezing}, taking as orientations $a=y, b=x$ (see the green dotted curve of Figure \ref{fig:OAT_linpar}). Related findings on the relevance of parity measurements to saturate the QFI throughout the evolution were reported in previous works \cite{PhysRevA.99.043621,PhysRevA.102.053315}, pointing out the crucial role played by parity conservation during the dynamics.

\section{Conclusions}
Certifying the resource content of a multipartite quantum system for metrology applications is an important task for quantum technologies. In this work, we introduced semidefinite programming algorithms to find the minimal sensitivity to unitary transformation, as quantified by the quantum Fisher information, compatible with an arbitrary set of mean values. These algorithms are general and improve over previous approaches by fully exploiting the convexity of the problem. We have shown that one can take advantage of symmetries to drastically reduce the numerical cost of the implementation. As illustrations, we have focused on relevant families of states of spin-$1/2$ ensembles. In the case of Dicke states, we have shown that a so-called Dicke squeezing parameter introduced in the past \cite{Zhang2014} overestimates the certifiable metrological usefulness. We have then focused on states generated in the one-axis-twisting dynamics \cite{Kitagawa1993}, where we obtained several significant results. First, we showed that using only the first and second moments of the collective spin, one can certify more metrological usefulness than given by the Wineland spin squeezing parameter \cite{winelandetal1994}. Including up to fourth-order moments, we made the surprising observation that, for up to a dozen of spins, the metrological power of so-called multi-headed cat states can be almost optimally certified. Finally, we found that using the sensitivity of parity operators (which are $N$-body correlation functions), close to the maximal QFI of the state can be certified throughout the whole dynamical evolution. Our results are directly relevant to many experiments where collective spin fluctuations are sampled and could be used to certify the metrological resource of such systems. Future applications of our approach could include spatially-structured systems, such as those studied in quantum simulators, and Appendix \ref{app_spin_chain} proposes a first numerical study in this direction. \\

\textbf{Code and data availability.--}  The code and data to construct the plots are available at the following link, \url{https://github.com/anubhavks/SDP_QFI_partialinfo.git}.

\acknowledgments{
We thank  Rafał Demkowicz-Dobrzański, Janek Kołodyński, Géza Tóth and Jordi Tura for useful discussions. IF acknowledges support from the European Union’s Horizon 2020 research and innovation programme under the Marie-Skłodowska-Curie grant agreement No 101031549 (QuoMoDys).
GMR, AKS, GRM, and ML acknowledge the support from: ERC AdG NOQIA; Ministerio de Ciencia y Innovation Agencia Estatal de Investigaciones (PGC2018-097027-B-I00/10.13039/501100011033, CEX2019-000910-S/10.13039/501100011033, Plan National FIDEUA PID2019-106901GB-I00, FPI, QUANTERA MAQS PCI2019-111828-2, QUANTERA DYNAMITE PCI2022-132919, Proyectos de I+D+I “Retos Colaboración” QUSPIN RTC2019-007196-7); MICIIN with funding from European Union NextGenerationEU(PRTR-C17.I1) and by Generalitat de Catalunya; Fundació Cellex; Fundació Mir-Puig; Generalitat de Catalunya (European Social Fund FEDER and CERCA program, AGAUR Grant No. 2021 SGR 01452, QuantumCAT \ U16-011424, co-funded by ERDF Operational Program of Catalonia 2014-2020); Barcelona Supercomputing Center MareNostrum (FI-2022-1-0042); EU (PASQuanS2.1, 101113690); EU Horizon 2020 FET-OPEN OPTOlogic (Grant No 899794); EU Horizon Europe Program (Grant Agreement 101080086 — NeQST), National Science Centre, Poland (Symfonia Grant No. 2016/20/W/ST4/00314); ICFO Internal “QuantumGaudi” project; European Union’s Horizon 2020 research and innovation program under the Marie-Skłodowska-Curie grant agreement No 101029393 (STREDCH) and No 847648 (“La Caixa” Junior Leaders fellowships ID100010434: LCF/BQ/PI19/11690013, LCF/BQ/PI20/11760031, LCF/BQ/PR20/11770012, LCF/BQ/PR21/11840013). AKS acknowledges support from the European Union’s Horizon 2020 Research and Innovation Programme under the Marie Skłodowska-Curie Grant Agreement No. 847517. SK acknowledges support from the National Science Center (Poland) grant No.2020/37/B/ST2/02134. Views and opinions expressed are, however, those of the author(s) only and do not necessarily reflect those of the European Union, European Commission, European Climate, Infrastructure and Environment Executive Agency (CINEA), nor any other granting authority. Neither the European Union nor any granting authority can be held responsible for them.
}
\bibliographystyle{plainnat}
\bibliography{biblio}

\onecolumn\newpage
\appendix

\section{Two SDP algorithms}
\label{sec:app_dtheta}

In this appendix, we detail the main technical considerations of two SDPs solving the central problem as formalized in Eq.~\eqref{eq:central_problem}. Firstly, for the algorithm based on maximizing fidelity as presented in the main text, we discuss the effect of the small parameter $\delta\theta$. Secondly, we provide a comprehensive derivation of the second SDP, which solves exactly the problem of Eq.~\eqref{eq:central_problem} by maximizing the variance of a locally unbiased estimator. Finally, we compare the performance of both algorithms in terms of accuracy and running computational times.

\subsection{Finite $\delta\theta$ considerations of the main-text SDP}

In order to illustrate the error introduced by a finite $\delta\theta$ in the SDP algorithm, we consider the following example: a system of $N=4$ qubits, in which we bound the QFI with respect to the generator $\hat{G}= \hat{J}_z$, imposing compatibility with the arbitrarily-chosen data: $\langle\hat{J}_x +0.2\hat{J}_z\rangle = 0.6078,\langle\hat{J}_y^2 \rangle = 0.5795$. As in the main text, we denote by $\hat{J}_a=\sum_{i=1}^4 \hat \sigma_i^a/2$ the collective spin along orientation $a$, with $\hat \sigma_i^a$ the Pauli matrices ($a\in\{x,y,z\}$). The SDP is solved for a range of values for $\delta\theta$. For each value of $\delta \theta$, we compute the QFI via two methods:

\begin{enumerate}
    \item By inverting the relation $\mathcal{F}_{\max} = 1-\mathcal{F}_Q[\hat{\rho}^*_{\delta\theta},\hat{G}](\delta\theta)^2 + {\cal O}(\delta\theta^4)$, the minimal QFI can be estimated from the fidelity resulting from the SDP, $\mathcal{F}_Q[\hat{\rho}^*_{\delta\theta},\hat{G}] = (1-\mathcal{F}_{\max})/(\delta\theta)^2$ (up to second-order corrections in $\delta\theta$).   
    \item By computing the QFI directly from the optimal $\hat{\rho}^*_{\delta\theta}$ which is itself obtained as an output of the SDP algorithm. With the spectral decomposition of the state $\hat{\rho} = \sum_{p}p\hat{\Pi}^{(p)} $, where $p$ are the eigenvalues and $\hat{\Pi}_p$ the projectors to the respective eigenspaces, the QFI with respect to $\hat{G}$ can be obtained with the closed formula,
    \begin{equation}
       \label{eq:QFI_formula}
        F_Q[\hat{\rho}, \hat{G}] = 2\sum_{p+p'>0}\frac{(p-p')^2}{(p + p')}\mathrm{Tr}(\hat{G}\hat{\Pi}^{(p)} \hat{G}\hat{\Pi}^{(p')}) \  \ .
    \end{equation}
    \label{eq:def_QFI}
    By construction, this method always gives a QFI realizable by a physical quantum state. 
\end{enumerate}

\begin{figure}[h]
  \centering
    \subfloat[\centering ]{{\includegraphics[width=0.45\textwidth]{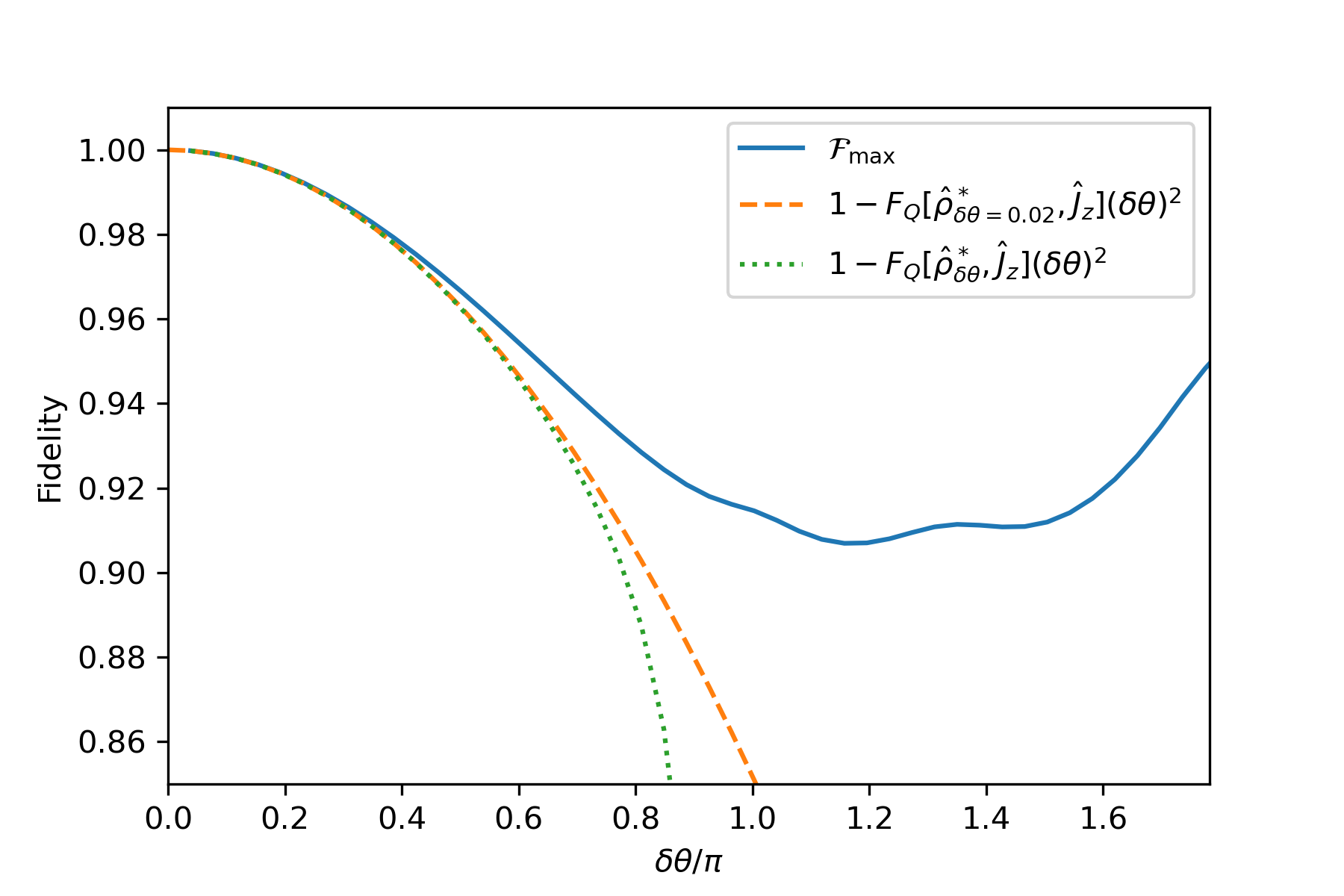} }}%
    \qquad
    \subfloat[\centering ]{{\includegraphics[width=0.45\textwidth]{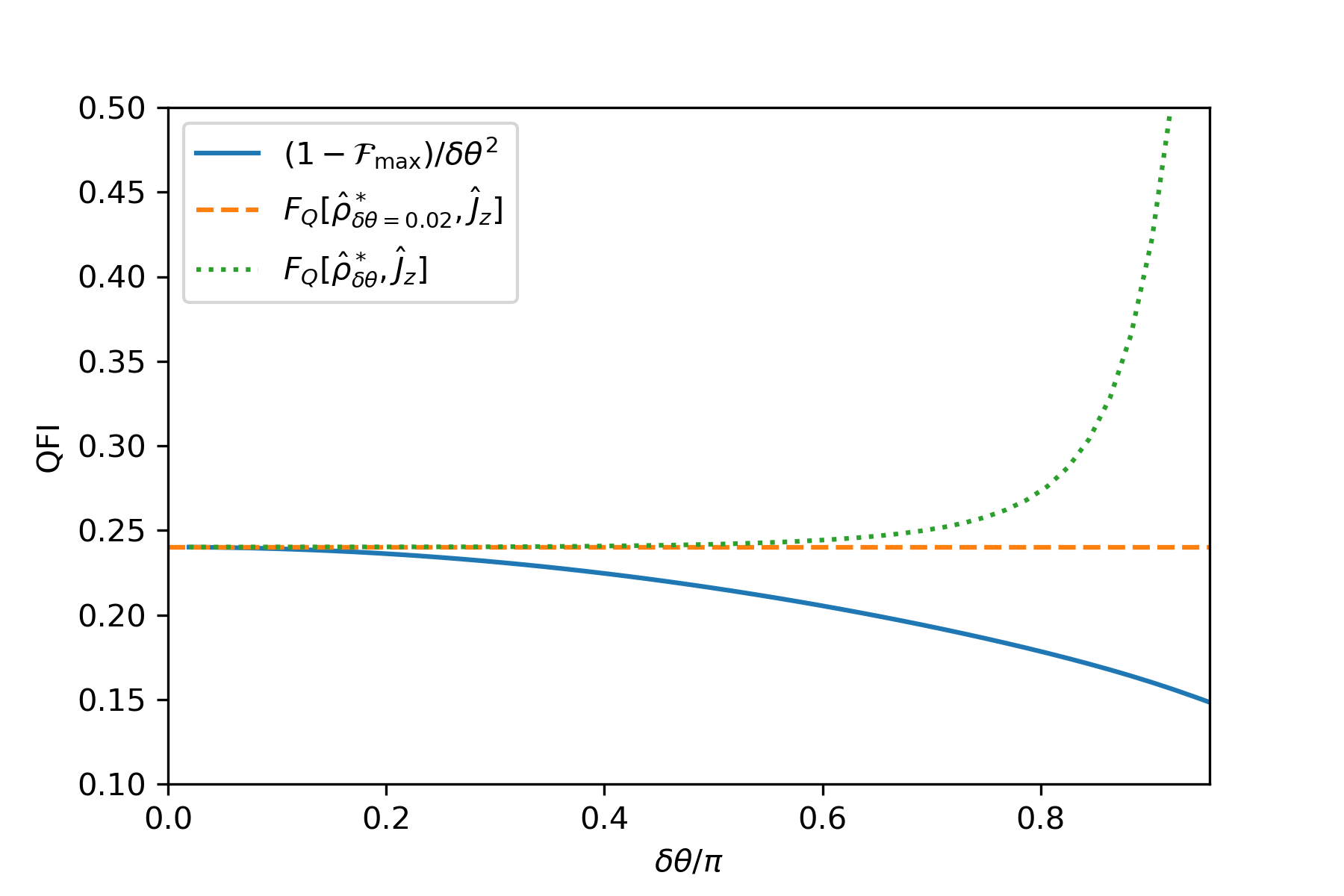} }}%
 \caption{Fidelity (left) and QFI (right) bounds as a function of $\delta\theta$. In blue (solid line), the value computed from the fidelity $\mathcal{F}_{\max}$ (SDP output). The orange dashed line corresponds to the final inferred QFI, $F_{Q}[\hat{\rho}^*_{\delta\theta = 0.02}, \hat{G}] = 0.2401$. The green dotted curve is deduced from the QFI of the optimal state $F_{Q}[\hat{\rho}^*_{\delta\theta}, \hat{G}]$ for each $\delta\theta$ using Eq.~\eqref{eq:QFI_formula}.}
 \label{fig:dtheta_error}
\end{figure}

We compare such ways to lower-bound the QFI in Figure~\ref{fig:dtheta_error}, where we plot both fidelity and QFI as a function of $\delta\theta$.  

From Figure \ref{fig:dtheta_error} we observe that for smaller values of $\delta\theta$ than a certain finite $\delta\theta_{\rm critical}$, the fidelity is a concave function of $\delta\theta$. After convergence of the SDP, this leads by means of method 1 to underestimation of the QFI bound. On the other hand, using method 2, a QFI value is given that it is exactly realizable by $\hat{\rho}^*_{\delta\theta}$. Thus, this method provides an overestimation, generally an upper bound to the method 1. In the limit $\delta\theta\rightarrow 0$, in practice $\delta\theta \sim 10^{-2}$, the two methods converge yielding a tight lower bound. Hereinafter, we will use method 2, because as shown in Figure~\ref{fig:dtheta_error}b it is the tightest in the limit of small $\delta\theta$.

\subsection{Alternative SDP algorithm }
In this section, we derive an alternative SDP algorithm computing the minimal QFI compatible with a set of mean values. This second algorithm does not involve the fidelity, and has the advantage that it is then no longer necessary to introduce the small parameter $\delta \theta$ as discussed in the previous section. Let us start with the following optimization problem whose solution is the inverse of the QFI \cite{helstrom1968minimum,zhou2023optimal} :
\begin{equation}
\begin{array}{crl}
\label{QFI_inv}
     F_Q [ \hat \rho, \hat G]^{-1}=&\min_{\hat{O} = \hat{O}^\dagger}& \mathrm{Tr}(\hat{\rho}\hat{O}^2)  \\
    &\mbox{s.t.}& \mathrm{Tr} \left( i [\hat \rho, \hat G] \hat O \right) = 1 \\
    &&\mathrm{Tr} \left( \hat \rho \hat O \right) = 0  \\
\end{array} \ .
\end{equation} 
The operator $\hat O$ can be interpreted as an observable representing the estimator for the unknown phase of the unitary transformation $\hat U(\theta)=e^{-i\theta \hat G}$. It can be decomposed as $\hat O = \sum_i \tilde \theta(i) \ket{i} \bra{i}$, where $\left\{\ket{i}\right\}
$ is the measurement basis and $\tilde \theta(i)$ is the value of the estimator of $\theta$ associated with the $i$-th measurement outcome. The objective function of the problem is the variance of the observable $\hat O$, while the constraints ensure local unbiasedness. The last constraint, $\mathrm{Tr} (\hat \rho \hat O) =0$, can be dropped without affecting the result of the minimization. Indeed, any $\hat O$  can be decomposed as $\hat O = \alpha \mathbb{1} + \hat O_0$, where $\mathrm{Tr}(\hat \rho \hat O_0) = 0$, changing $\alpha$ does not affect the first constraint, and $
\mathrm{Tr} (\hat \rho \hat O^2) = \mathrm{Tr} (\hat \rho \hat O_0^2) + \alpha^2 \mathrm{Tr}(\mathbb{1})$ is clearly minimal when $\alpha =0$, which implies that $\mathrm{Tr} (\hat \rho \hat O) = 0$. Let us reformulate as an SDP the former problem:
\begin{equation}
\label{QFI_inv_SDP}
\begin{array}{crl}
     F_Q [ \hat \rho, \hat G]^{-1}=&\min_{\hat{O} , \hat O_2}& \mathrm{Tr}(\hat{\rho}\hat{O}_2)  \\
    &\mbox{s.t.}& \mathrm{Tr} \left( i [\hat \rho, \hat G] \hat O \right) = 1 \\
    &&\begin{pmatrix}
    \hat{O}_2 & \hat{O} \\
    \hat{O} & \mathbb{1}
    \end{pmatrix} \succeq 0  \\
\end{array} \ .
\end{equation} 
As a consequence of the last constraint, matrices $\hat O$ and $\hat O_2$ are Hermitian, and $\hat O_2 \succeq \hat O^2$ due to Schur's complement condition. Therefore, the minimum of the objective function is achieved when $\hat O_2 = \hat O^2$ (because $\hat \rho$ is positive-semidefinite), which means that problems Eq.~\eqref{QFI_inv} and Eq.~\eqref{QFI_inv_SDP} are equivalent. Our task is to minimize the QFI over $ \hat \rho$ satisfying certain constraints, so the inverse of the QFI must be maximized and the problem we want to solve is:
\begin{equation}
\label{maxmin}
    \begin{array}{crl}
     \max{\hat \rho} &\min_{\hat{O} , \hat O_2}& \mathrm{Tr}(\hat{\rho}\hat{O}_2)  \\
    &\mbox{s.t.}& \mathrm{Tr} \left( i [\hat \rho, \hat G] \hat O \right) = 1 \\
    &&\begin{pmatrix}
    \hat{O}_2 & \hat{O} \\
    \hat{O} & \mathbb{1}
    \end{pmatrix} \succeq 0  \\
    &&\mathrm{Tr}
  (\hat{\mathbf{ D}}\hat{\rho}) = \langle\hat{\mathbf{D}}\rangle \\
    &&\hat{\rho} \succeq 0
\end{array} \ .
\end{equation}
To simplify the problem, we are going to change it from ``max-min'' to ``max-max'' -- to achieve this, we will use a strong duality theorem for SDPs, and change the minimization over $\hat O, \hat O_2$ to a maximization. 

Let us start with rewriting the minimization problem of Eq.~\eqref{QFI_inv_SDP} into a canonical form:
\begin{equation}
\label{SDP_canon_min}
    \begin{array}{crl}
     &\min_{\hat{X}}& \mathrm{Tr}(\hat{C}\hat{X})  \\
    &\mbox{s.t.}& \Phi(\hat{X}) = \hat B\\
    && \hat X \succeq 0  \\
\end{array} \ ,
\end{equation}
where $\Phi $ is a linear, Hermitian-preserving map (notice that $\hat X$ and $\hat B$ are not necessarily of the same dimension).
The dual maximization problem, whose solution is the same as of the primal one, reads \cite{watrous2018theory}
\begin{equation}
\label{SDP_canon_dual}
\begin{array}{crl}
     &\max_{ \hat Y = \hat Y^\dagger}& \mathrm{Tr} (\hat B \hat Y)  \\
    &\mbox{s.t.}& \Phi^*(\hat Y) \preceq \hat C \\
\end{array} \ ,
\end{equation}
where $\Phi^*$ is the adjoint map, satisfying 
\begin{equation}
\label{adjoint_map}
\mathrm{Tr} \left( \Phi(\hat X) \hat Y \right) = \mathrm{Tr} \left( \hat X \Phi^*(\hat Y) \right) 
\end{equation} 
for all Hermitian $\hat X, \hat Y$. Let $d$ be the dimension of $\hat \rho$, and $\Phi$ be a  map from $2d \times 2d$ to $(2d+1) \times (2d+1)$ Hermitian matrices, defined as
\begin{equation}
    \label{Phidef}
    \Phi \left(\hat X = \begin{pmatrix} \hat X_{00} & \hat X_{01} \\ \hat X_{01}^\dagger & \hat X_{11} \end{pmatrix} \right) = \begin{pmatrix} \frac{1}{2} \mathrm{Tr} \left( i [\hat \rho, \hat G] (\hat X_{01} + \hat X_{01}^\dagger)\right) & 0 & 0 \\ 0 & i(\hat X_{01} - \hat X_{01}^\dagger) & 0_{d \times d} \\ 0 & 0_{d \times d} & \hat X_{11} \end{pmatrix},
\end{equation}
$0_{d \times d}$ is a $d \times d$ zero matrix, $\hat X_{ij}$ are $d \times d$ blocks of $\hat X$.
The problem Eq.~\eqref{QFI_inv_SDP} is equivalent to Eq.~\eqref{SDP_canon_min}  for such $\Phi $ if
    \begin{equation}
    \label{Cdef}
        \hat C = \begin{pmatrix}
    \hat \rho & 0_{d \times d} \\
    0_{d \times d} & 0_{d \times d}
    \end{pmatrix},
    \end{equation}
    and  

    \begin{equation}
    \label{Bdef}
        \hat B =\begin{pmatrix} 1 & 0 &0 \\ 0 & 0_{d \times d} & 0_{d \times d} \\ 0 & 0_{d \times d} & \mathbb{1}_{d \times d} \end{pmatrix}.
    \end{equation}

Since $\hat X$ is constrained to be positive-semidefinite, it must also be Hermitian, which, together with equalities $\hat X_{01} = \hat X_{01}^\dagger$ and $\hat X_{11} = \mathbb{1}$ (which follow from condition $\Phi(\hat X) = \hat B$) implies that $\hat X$ has structure $\hat X = \begin{pmatrix} \hat O_2 & \hat O \\ \hat O & \mathbb{1}\end{pmatrix}$, where $\hat O$, $\hat O_2$ are both Hermitian. Moreover, $\mathrm{Tr} \left( \hat X \hat C \right) = \mathrm{Tr} \left( \hat \rho \hat O_2 \right) $, and from $\Phi(\hat X) = \hat B$ follows that $\mathrm{Tr} \left( i [\hat \rho, \hat G] \hat O \right) = 1$, which shows that the problem Eq.~\eqref{QFI_inv_SDP} is indeed equivalent to the canonical form Eq.~\eqref{SDP_canon_min} with parameters given by Eq.~\eqref{Phidef}, Eq.~\eqref{Cdef}, Eq.~\eqref{Bdef}. Let us write down the Hermitian matrix $\hat Y$ as 
\begin{equation}
\hat Y = 
    \begin{pmatrix}
        y_{00} & y_{01} & y_{02} \\
        y_{01}^* & \hat Y_{11} & \hat Y_{12} \\
        y_{02}^* & \hat Y_{12}^\dagger& \hat{Y}_{22} 
    \end{pmatrix},
\end{equation}
where $\hat{Y}_{ij}$ are $d \times d $ blocks, and plug in this representation of $\hat Y$ into Eq.~\eqref{adjoint_map} with $\Phi$ defined by Eq.~\eqref{Phidef}. After expanding both sides of Eq.~\eqref{adjoint_map} and comparing them, we obtain
\begin{equation}
    \Phi^*(Y) = \begin{pmatrix}
        0_{d \times d} & i \hat Y_{11}+ \frac{i}{2} y_{00} [\hat \rho, \hat G] \\
        -i \hat Y_{11} +\frac{i}{2} y_{00} [\hat \rho, \hat G] &  \hat Y_{22}
    \end{pmatrix}.
\end{equation}
Moreover, $\mathrm{Tr} (\hat B \hat Y) = y_{00} + \mathrm{Tr} (\hat{Y}_{22})$, so the SDP problem dual to Eq.~\eqref{QFI_inv_SDP} is
\begin{equation}
\label{dual_QFI_inv}
    \begin{array}{crl}
     F_Q [ \hat \rho, \hat G]^{-1}=&\max_{y_{00}  , \hat Y_{11}, \hat Y_{22} }& y_{00} + \mathrm{Tr}(\hat{Y_{22}})  \\
    &\mbox{s.t.}& \begin{pmatrix}
    \hat{\rho} & -i \hat Y_{11}- \frac{i}{2} y_{00} [\hat \rho, \hat G] \\
    i \hat Y_{11}- \frac{i}{2} y_{00} [\hat \rho, \hat G] & - \hat{Y}_{22}
    \end{pmatrix} \succeq 0 \\
\end{array}, \
\end{equation}

$y_{00}$ is real, and $\hat Y_{11}, \hat Y_{22}$ are Hermitian because $\hat{Y}$ is Hermitian. We are now almost ready to express problem Eq.~\eqref{maxmin} as a single SDP, in which the maximization is performed also over  $\hat \rho$. The only problem is that variables $y_{00}$ and $\hat \rho$ are multiplied in the constraint of Eq.~\eqref{dual_QFI_inv}. To deal with this issue, let us make the following observation: when block matrix satisfies $\begin{pmatrix} A_{00} & A_{01} \\ A_{10} & A_{11} \end{pmatrix} \succeq 0$, then $\begin{pmatrix} x^2 A_{00} & x A_{01} \\ x A_{10} & A_{11} \end{pmatrix} \succeq 0$ for any real $x$. This is a direct consequence of Schur's complement condition for matrix positivity. This observation allows us to change the constraint from  Eq.~\eqref{dual_QFI_inv} to 

\begin{equation}
\label{constr_rhop}
    \begin{pmatrix}
    y_{00}^2 \hat{\rho} & -i \hat M- \frac{i}{2}  [y_{00}^2 \hat \rho, \hat G] \\
    i \hat M- \frac{i}{2}  [y_{00}^2 \hat \rho, \hat G] & - \hat{Y}_{22}
    \end{pmatrix} \succeq 0,
\end{equation}
 where $\hat M = y_{00} \hat Y_{11}$ is an arbitrary Hermitian matrix, which will be our new independent variable (notice that neither $\hat Y_{11}$ nor $\hat M$ enter into the problem's objective). Let us introduce a new matrix variable $\hat \rho' = y_{00}^2 \hat \rho$ satisfying $\mathrm{Tr}(\hat \rho ') = y_{00}^2$, and impose a condition
 \begin{equation}
     \begin{pmatrix}
         \mathrm{Tr}(\hat \rho') & y_{00} \\
         y_{00} & 1
     \end{pmatrix} \succeq 0,
 \end{equation}
 which is equivalent to $\mathrm{Tr}(\hat \rho') \ge y_{00}^2$. When the objective function $y_{00}^2 + \mathrm{Tr}(\hat Y_{22})$ is maximized, then the above inequality becomes equality. This allows us to treat $y_{00}$ and $\hat \rho '$ as independent variables. Importantly, constraint Eq.~\eqref{constr_rhop} is linear with $ \hat \rho'$. Finally, the constraints $ \mathrm{Tr} ( \hat{\bm{D}} \hat \rho ) = \langle \hat{\bm{D}} \rangle $ are equivalent to $\mathrm{Tr} (\hat{\bm{D}} \hat \rho' ) = \langle \hat{\bm{ D}} \rangle \mathrm{Tr} (\hat \rho')$. All these observations allow us to write the expression for the inverse of minimal QFI,
 \begin{equation}
     \begin{array}{crl}
     [F_Q(\hat{G})]^{-1}=&\max_{ \hat \rho }& F_Q^{-1} [\hat \rho, \hat G]  \\
    &\mbox{s.t.}& \mathrm{Tr} ( \hat{\bm{D}} \hat \rho ) = \langle \hat{\bm{D}} \rangle \\
    && \rho \succeq 0 \\
\end{array} \ ,
 \end{equation}
 as a single SDP:
\begin{equation}
\label{eq:SDP_stas}
\begin{array}{crl}
     [F_Q(\hat{G})]^{-1}=&\max_{ \hat \rho', y_{00}, \hat{M}, \hat{Y}_{22}}& y_{00} + \mathrm{Tr} (\hat{Y}_{22}) \\
    &\mbox{s.t.}& \begin{pmatrix}
     \hat{\rho}' & -i \hat M- \frac{i}{2}  [ \hat \rho', \hat G] \\
    i \hat M- \frac{i}{2}  [ \hat \rho', \hat G] & - \hat{Y}_{22}
    \end{pmatrix} \succeq 0 \\
    &&\begin{pmatrix}
         \mathrm{Tr}(\rho') & y_{00} \\
         y_{00} & 1
     \end{pmatrix} \succeq 0 \\
     && \mathrm{Tr} (\hat{\bm{D}} \hat \rho' ) = \langle \hat{\bm{ D}} \rangle \mathrm{Tr} (\hat \rho') \\
     &&  \hat{Y}_{22} = \hat{Y}_{22}^\dagger, \hat M = \hat M^\dagger
\end{array} .\ 
\end{equation}
\subsection{Comparison}
\begin{figure}[h]
  \centering
    \subfloat[\centering ]{{\includegraphics[width=0.45\textwidth]{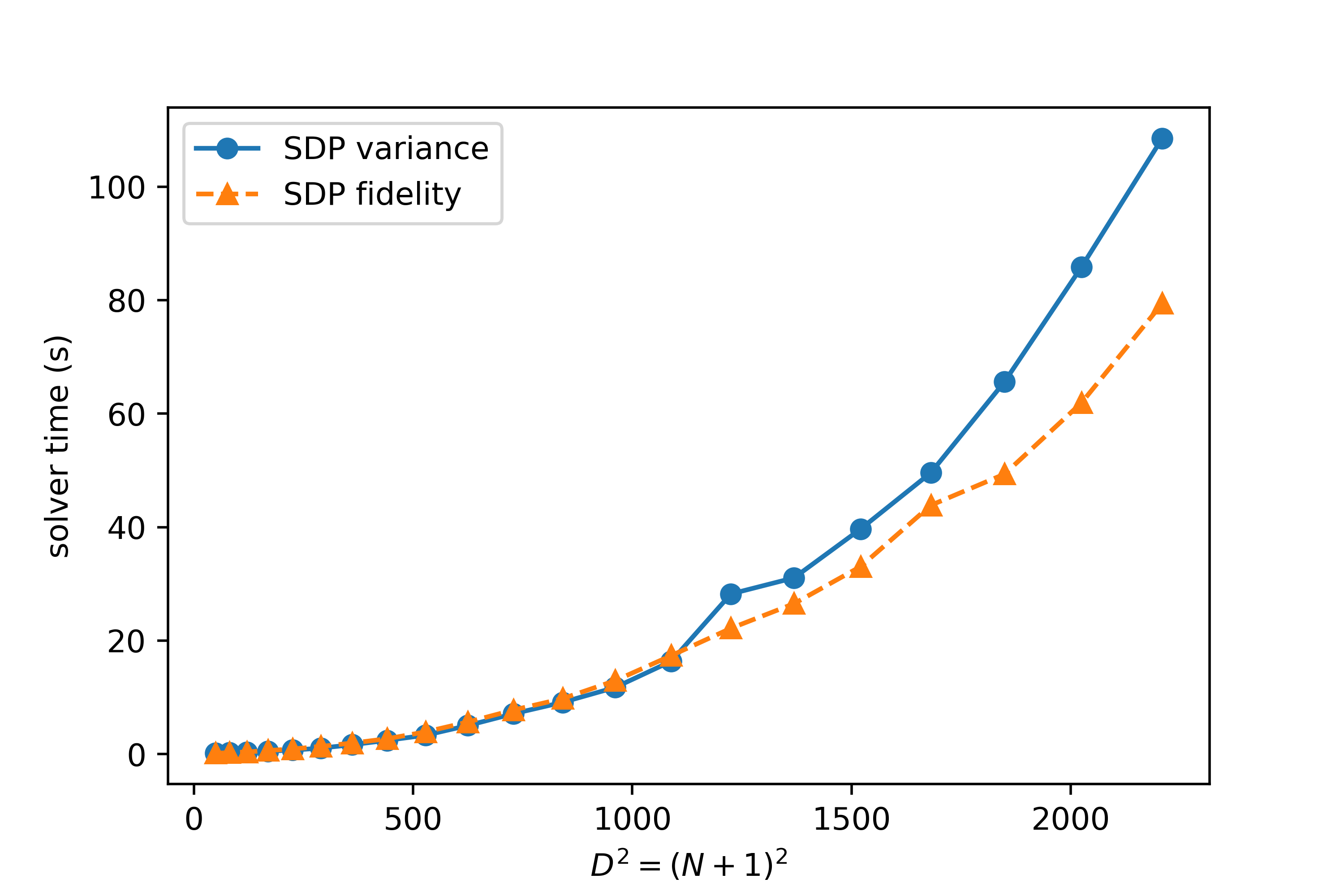} }}%
    \qquad
    \subfloat[\centering ]{{\includegraphics[width=0.45\textwidth]{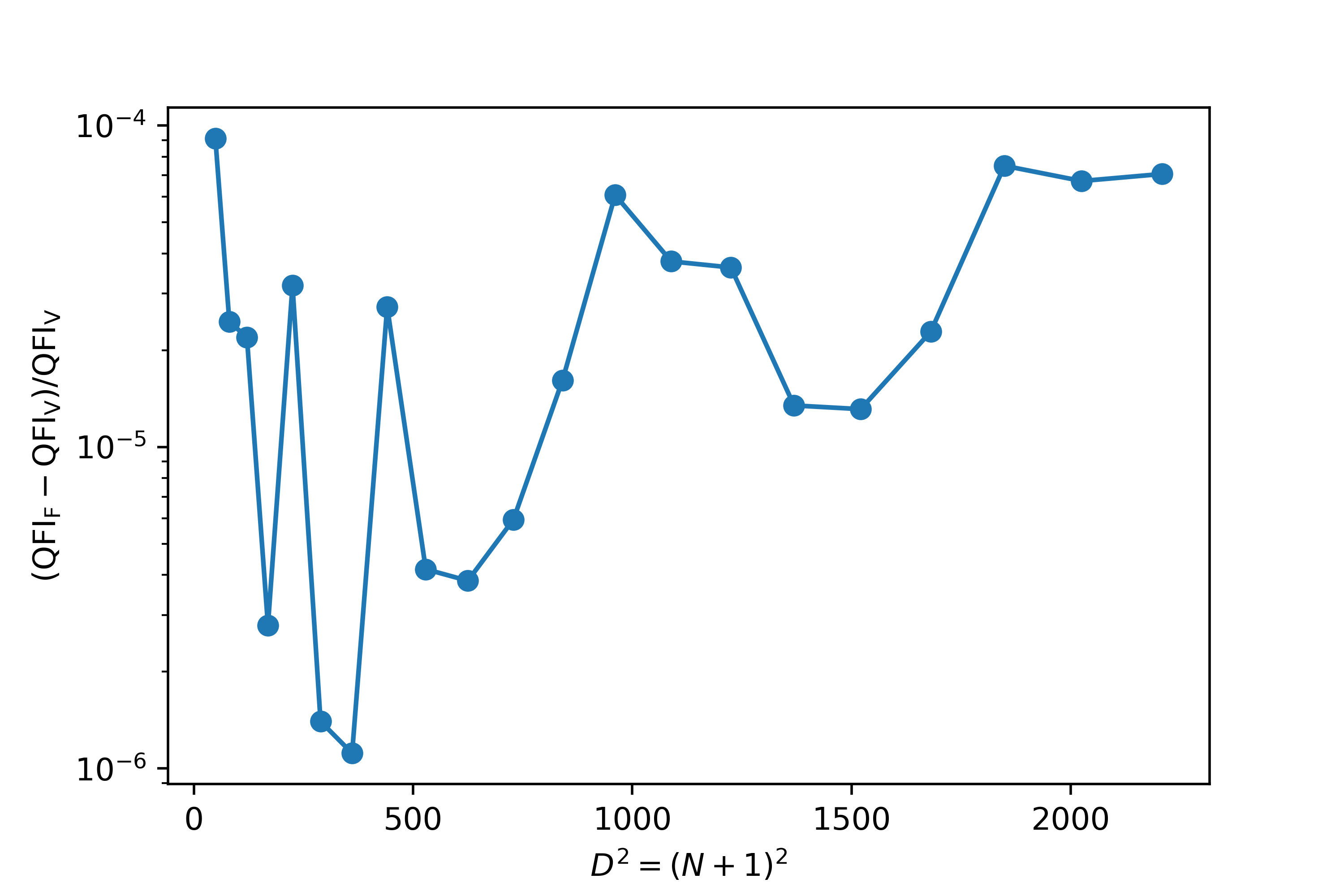} }}%
 \caption{a) Computing time for both approaches for different values of $N$ and $\delta\theta = 0.01$ for the SDP used in the main text. b) Relative difference of the two bounds ($F$ stands for fidelity and $V$ for variance) for the same values of $N$.}
 \label{fig:comparison}
\end{figure}

In this section, we compare the efficiency and accuracy of the two previously-introduced SDP algorithms. As an example, we consider a system of $N$ qubits in a totally symmetric state, i.e. $\langle \hat{\mathbf{J}}^2\rangle = (N/2)(N/2+1)$, where $\hat{\mathbf{J}}^2$ is the total spin. In the data, we include a mean spin $\langle\hat{J}_x \rangle = 0.8$ and the fluctuations in an orthogonal direction $\langle\hat{J}_y^2 \rangle = 0.4$. We find the minimal QFI with respect to the generator $\hat{J}_z$ compatible with such expectation values via the two methods, varying $N$ between $N=4$ and $N = 46$ (for even values). For each $N$ we record both the computing time and corresponding QFI lower-bound. The number of particles $N$ is related to the dimension of the totally symmetric subspace as $D= N+1$. The corresponding results are shown in Figure \ref{fig:comparison}.

In Figure \ref{fig:comparison}, we observe how the fidelity-based SDP is marginally more efficient than the variance-based SDP. This fact stems from the extra parameter and constraint existing in the latter approach.  
In panel b) we study the discrepancy between the two bounds. We find that due to the finite $\delta\theta$ the fidelity-based method slightly overestimates the variance bound, within a $10^{-4}$ relative accuracy.

\section{Permutation invariance}
\label{sec:app_symmetries}
In this section, we discuss the implementation of symmetries in order to reduce the computational complexity of our SDP algorithm. For the sake of concreteness, we consider a system of $N$ qubits. A generic density matrix describing the system is fully specified by exponentially-many parameters. We will show that if the observables and the generator are invariant under permutations of the qubits (PI), it is sufficient to optimize over $\mathcal{D}_N = (1+\lfloor N/2 \rfloor)(3(1+N)^2-2(2+3N)\lfloor N/2 \rfloor+4\lfloor N/2 \rfloor^2)/3-1$ (that is: polynomially-many) real parameters representing a state which is block-diagonal with respect to the symmetry sectors. We hence obtain a drastic scaling reduction.

We first define the local basis. The state of each spin $i\in \{1,2,..,N\}$ is fully described by the Pauli vector $\hat{\boldsymbol{\sigma}}^{(i)}=(\hat{\sigma}^{(i)}_x, \hat{\sigma}^{(i)}_y, \hat{\sigma}^{(i)}_z )$, where $\hat{\sigma}^{(i)}_a$ applies Pauli matrix $a$ in party $i$ with trivial action in the remaining qubits. From these operators, typical PI observables that we can form are collective spin $\hat{\mathbf{J}}= \sum_{i=1}^N\mathbf{S}^{(i)} $ (elementary), where $\mathbf{S} = \hat{\boldsymbol{\sigma}}/2$, and moments, i.e., polynomials thereof. All those operators have in common that they commute with a unitary representation of any permutation of $N$ parties, $\pi\in \mathcal{S}_N$,
$\hat{U}_\pi$. Such unitary is described by its action in the computational basis $\hat{U}_\pi\ket{a_1b_2c_2...} = \ket{a_{\pi(1)}b_{\pi(2)}c_{\pi(3)}...}$. \\

\noindent\textbf{Sufficiency of optimizing over PI states.--} In this paragraph we show that it is sufficient to optimize over PI states if both the observables in $\hat{\mathbf{D}}$ and the generator $\hat{G}$ are PI. The proof consists in two steps. First, we notice that if there is a state compatible with the data, then there always exists also a PI state compatible with the data. Indeed, for any state $\hat{\rho}$ we can generate the PI state by applying the averaging map  $\hat{\rho}_{\rm PI} = (N!)^{-1}\sum_{\pi\in \mathcal{S}_N}\hat{U}_\pi^\dagger\hat{\rho}\hat{U}_\pi$. Both states $\hat{\rho}$ and $\hat{\rho}_{\rm PI}$ will yield the same expectation values for PI observables. The second step is to establish a relation between the QFI of $\hat{\rho}$ and that of $\hat{\rho}_{\rm PI}$. Since  $\hat{U}_{\rm{PI}}$ commute with the generator $\hat{G}$, $F_Q[\hat{\rho}, \hat{G}] = F_Q[\hat{U}_\pi^\dagger\hat{\rho}\hat{U}_\pi, \hat{G}]$ for any $\pi\in \hat{S}_N$. Finally, by the convexity of QFI from the averaging map, we see that $F_Q[\hat{\rho}_{\rm PI}, \hat{G}]\leq F_Q[\hat{\rho}, \hat{G}]$, concluding that in our optimization for the search of the minimum QFI, PI states are sufficient.    \\

\noindent \textbf{Block-diagonal representation of PI states.--} Here, we derive an efficient representation of PI states. First, one can check that any elementary PI operator $\hat{A}_{\rm PI}$ commutes with the total spin $\hat{\mathbf{J}}^2$. Hence, it can be decomposed in the block-diagonal form, $\hat{A}_{\rm PI} = \oplus_{J=J_{\min}}^{J_{\max}}\hat{A}^{(J)}$, where $J$ label the degenerate eigenspaces of $\hat{\mathbf{J}}^2$ with eigenvalue $J(J+1)$, where $J_{\min} = 0,1/2$,  for even, odd $N$ and $J_{\max} = N/2$. All sectors $J$ have coarse-grained dimension $2J+1$, which is the number of PI inequivalent ways to combine $N$ spins-$1/2$ to form a collective state of total spin $J$. Now, each vector represents a set of highly degenerated orthogonal states that cannot be distinguished by any PI observable. With previous considerations, we can identify each sector $J$ as a magnified spin, of length $J$. Accordingly, the irreducible representations are $\hat{\mathcal{J}} = \oplus_{J=J_{\mathrm{min}}}^{J_{\max}}\hat{\mathbf{j}}^{(J)}$ with $\hat{\mathbf{j}}^{(J)}$ the vector of spin-$J$ matrices. We can establish the correspondence $\mathbf{v} \cdot\hat{\mathcal{J}}\sim  \mathbf{v}\cdot\hat{\mathbf{J}}$ as they fulfill the same algebra. Moreover, beyond linear combination, for any polynomial function $f$, $f(\hat{\mathcal{J}}) \sim  f(\hat{\mathbf{J}})$. \\

We can finally count the number of real parameters of our representation. Each block $J$ is of dimension $2J+1=D_J$, representing generally an $D_J\times D_J$ Hermitian matrix. An Hermitian matrix of size $D_J$ is described by $D_J^2$ real parameters. Adding all blocks yields $\sum_{J = J_{\min}}^{N/2}D_J^2 = \mathcal{D}_N +1$. Note that we can remove a parameter due to the unit trace constraint. \\

With the above block structure, the SDP based on maximizing fidelity, Eq.~\eqref{eq:sdp_FQ} of the main text, can be expressed in the following way:
\begin{equation}
   \label{eq:sdp_FQ_symm}
   \begin{array}{crl}
     \sqrt{{\cal F}_{\rm max}} =&\max_{\left\{\substack{\hat{\varrho}^{(J)} = [\hat{\varrho}^{(J)}]^\dagger\\\hat{L}^{(J)}}\right\}_{J=J_{\min}}^{J_{\max}}}& \sum_{J=J_{\min}}^{J_{\max}}\Re{\mathrm{Tr}(\hat{L}^{(J)})}  \\
    &\mbox{s.t.}&  \forall J\in \{J_{\min},J_{\min}+1,...,J_{\max} \} \\
    &&\begin{pmatrix}
   \hat{\rho}_+^{(J)}   & [\hat{L}^{(J)}]^\dagger \\
    \hat{L}^{(J)} & \hat{\rho}_-^{(J)} 
    \end{pmatrix} \succeq 0   \\
    & &\hat{\rho}_\pm ^{(J)} =  \hat{\varrho}^{(J)} \pm  i \delta\theta [\hat{G}^{(J)},\hat{\varrho}^{(J)}]  \ , \\
    & & \sum_{J=J_{\min}}^{J_{\max}}\mathrm{Tr}(\hat{\mathbf{D}}^{(J)}\hat{\varrho}^{(J)})= \langle\hat{\mathbf{D}} \rangle
     \end{array} \ .
\end{equation} 
Our second approach, as introduced in Appendix \ref{sec:app_dtheta}.2 Eq.~\eqref{eq:SDP_stas}, also admits a similar block decomposition:
\begin{equation}
\label{eq:SDP_stas_block}
\begin{array}{crl}
[F_Q(\hat G)]^{-1}=&\max_{\left\{\substack{y_{00}^{(J)} \in \mathbb{R} \\\hat{\rho}'^{(J)} =[\hat{\rho}'^{(J)}]^\dagger \\ \hat{Y}_{22}^{(J)} = [\hat{Y}_{22}^{(J)}]^\dagger \\ \hat{M}^{(J)} = [\hat{M}^{(J)}]^\dagger}\right\}_{J=J_{\min}}^{J_{\max}}}& \sum_{J=J_{\min}}^{J_{\max}}\left[y_{00}^{(J)} + \mathrm{Tr}(\hat{Y}_{22}^{(J)})\right] \\
    &\mbox{s.t.}&  \forall J\in \{J_{\min},J_{\min}+1,...,J_{\max} \} \\
 && \begin{pmatrix}
     \hat{\rho'}^{(J)} & -i \hat{M}^{(J)} - i[\hat{G}^{(J)},\hat{\rho}'^{(J)}]/2\\
     i \hat{M}^{(J)}-i[\hat{G}^{(J)},\hat{\rho}'^{(J)}]/2 & -\hat{Y}_{22}^{(J)}
 \end{pmatrix}\succeq 0\\
&& \begin{pmatrix}
\mathrm{Tr}(\hat{\rho}'^{(J)}) & y_{00}^{(J)} \\
y_{00}^{(J)} & 1 
\end{pmatrix} \succeq 0 \ , \\
&& \sum_{J=J_{\min}}^{J_{\max}}\mathrm{Tr}(\hat{\mathbf{D}}^{(J)}\hat{\rho}'^{(J)}) = \langle \hat{\mathbf{D}}\rangle \sum_{J=J_{\min}}^{J_{\max}}\mathrm{Tr}(\hat{\rho}'^{(J)})
\end{array} \ .
\end{equation} 
Note that all the sectors must be taken into account for the optimization. Minimization over a single sector will lead generally to an upper bound of the true minimum. Besides, more than one sector may be required to meet the data constraints.  However, if we set the total spin to its extremal value, $\langle\hat{\mathbf{J}}^2 \rangle = J_{\max}(J_{\max} +1)$, optimizing over the total symmetric subspace $J = J_{\max} =  N/2$ is sufficient.

\section{Comparison with alternative approaches}
\label{sec:app_apellaniz}

The central problem of the paper, Eq.~\eqref{eq:central_problem}, has appeared in several contexts throughout the literature. In this appendix, for completeness, we provide a concise overview of some of its manifestations and how the corresponding solutions compare with the present approach.   

\subsection{Relaxations}

Here, we describe the methods that provide a valid lower bound for our problem, but however, there is no guarantee of its optimality, nor of the existence of a quantum realization for it. Note that this is not the case for our SDP-based approach, as by construction, there is a quantum state compatible with the data with minimal QFI. \\

\noindent\textbf{Loong Len et al.--} In Ref. \cite{Len2022} a method was proposed to lower bound the classical Fisher information (FI), $F$, of an observable $\hat{O}$ from moments of it. Up to K-th moments, they defined:

\begin{equation}
    F^{(K)} = \mathbf{b}^TA^{-1}\mathbf{b}\leq F \ ,
\end{equation}
where the vector $\mathbf{b}$ has components $b_p = \partial_\theta\langle \hat{O}^p\rangle =\langle i[\hat{G},\hat{O}^p]\rangle$, and the matrix $A$, $A_{pq} = \langle\hat{O}^{p+q} \rangle$ for $p,q\in\{0,1,..,K\}$. The sensitivity $\Xi_O^{-2}$ is recovered in the first order ($K=1$). By construction, $F^{(K)}\leq F^{(K+1)}$, and in the limit $K\rightarrow \infty$, it converges to $F$. The FI is a lower bound of the QFI. Thus, $F^{(K)}$ constitutes also a proper lower bound of QFI. Such values can be compared in our approach with the minimal QFI compatible with the data $\langle\hat{\mathbf{D}}\rangle = \{\langle i[\hat{G},\hat{O}^p] \rangle, \langle\hat{O}^{p+q} \rangle \}_{p,q\in\{0,1,..,K \}}$.   \\

\noindent\textbf{Gessner et al.--} In Ref. \cite{Gessner2019} they ask which generator $\hat{G}\in \mathrm{Span}(\hat{\mathbf{G}})$ and observable $\hat{O}\in \mathrm{Span}(\hat{\mathbf{G}} \oplus \hat{\mathbf{O}})$ capture the maximal sensitivity $\Xi^{-2}_{O}$, which reads:

\begin{equation}
\Xi_{O}^{-2} = \max_{\mathbf{n}\ \mathrm{s.t.}||\mathbf{n}|| = 1}\mathbf{n}^T\underbrace{P_{\hat{\mathbf{G}}}\Im{\mathcal{C}} \Re{\mathcal{C}}^{-1} \Im{\mathcal{C}} P_{\hat{\mathbf{G}}}}_{:= M}\mathbf{n} = \lambda_{\max}(M) \ ,
\end{equation}
where $P_{\hat{\mathbf{G}}}$ projects to the subspace generated by ${\hat{\mathbf{G}}}$ and $\mathcal{C} = \langle\hat{\mathbf{O}}\hat{\mathbf{O}}^T\rangle -\langle\hat{\mathbf{O}} \rangle\langle\hat{\mathbf{O}} \rangle^T$ is the complex covariance matrix. The optimal generator is $G =\mathbf{n}_{G}\cdot\hat{\mathbf{G}}$ where $\mathbf{n}_G$ is the maximal eigenvector of $M$ and the observable is $\hat{O} = \mathbf{n}_O\cdot( \hat{\mathbf{O}}\oplus \hat{\mathbf{G}})$, with $\mathbf{n}_O=\Re{\mathcal{C}}^{-1}\Im{\mathcal{C}}\mathbf{n}_G/||\mathbf{n}_O||$ yielding $\Xi_{O}^{-2} = \langle i[\hat{G},\hat{O}] \rangle^2/(\Delta\hat{O})^2 = \lambda_{\max}(M)$. 

This approach is used in the main text to find the optimal generator for a set of expectation values during OAT dynamics. 
For example, if spin and its fluctuations are given, the method recovers the squeezing and anti-squeezing axes for $\mathbf{n}_O$ and $\mathbf{n}_G$ respectively. By taking $\hat{\mathbf{O}}$ up to second moments, we can derive a nonlinear squeezing parameter beyond spin. Such bound is also compared with our SDP value in the main text with the same data, i.e. considering data $\langle\hat{\mathbf{D}}\rangle$ up to the fourth moments of the collective spin. In general, this approach yields a valid but not optimal lower bound to the QFI given the data.  \\

\noindent\textbf{Tóth et al.--} In Ref. \cite{Toth2015} an SDP-based approach was proposed to find lower bounds on quantities defined as convex roofs of polynomials of operator expectation values. It can also find a lower bound of the QFI given few arbitrary expectation values. The starting point is to notice that, as proven in Ref. \cite{Yu2013}, the QFI is the convex roof of the variance:
\begin{equation}
    \label{eq:convex_roof}
    F_Q[\hat{\rho},\hat{G}] =  4\min_{\hat{\rho} = \sum_kp_k\ketbra{\psi_k}} \sum_k p_k(\Delta\hat{G})^2_{\ket{\psi_k}} \ .
\end{equation}
Then, the variance of a pure state $\ket{\psi}\in\mathcal{H}$ can be linearized in the two-copy space, $(\Delta \hat{G})^2_{\ket{\psi}} = \mathrm{Tr}[(\hat{G}^2\otimes \mathbb{I} - \hat{G}\otimes \hat{G})(\ketbra{\psi}\otimes \ketbra{\psi})]$. In this embedding, Eq.~\eqref{eq:convex_roof} is rewritten as a minimization of separable symmetric states $\hat{\sigma}\in \mathrm{SEP}(\mathcal{H}\otimes\mathcal{H})$. The separability condition in general cannot be incorporated as a positive-semidefinite constraint. However, one can relax the previous restriction to states with positive partial trace (PPT) and formalize the following SDP: 

\begin{equation}
\label{eq:SDP_Geza}
    \begin{array}{crl}
   \mathrm{F}_Q(\hat{G})\geq  &4\min_{\hat{\sigma} = \hat{\sigma}^\dagger, \hat{\sigma}\succeq 0}& \mathrm{Tr}[(\hat{G}^2\otimes \mathbb{I} - \hat{G}\otimes \hat{G})\hat{\sigma}] \\
    &\mbox{s.t.}&  \hat{S}\hat{\sigma} =\hat{\sigma} \hat{S} = \hat\sigma,\  \hat{\sigma}^{t}\succeq 0  \\
    & & \mathrm{Tr}[\hat{\mathbf{D}}\mathrm{tr}(\hat{\sigma})] = \langle\hat{\mathbf{D}} \rangle
    \end{array} \ ,
\end{equation}
where $\hat{S}$ is the exchange operator (swapping the two subsystems) with action $\hat{S} \ket{\psi}\otimes \ket{\phi} = \ket{\phi}\otimes \ket{\psi}$, $t$ the partial transpose and $\mathrm{tr}$ the partial trace with respect to a subsystem. Note that this method requires the optimization over a matrix of size $D^2$ (with $D$ the dimension of the Hilbert space $\mathcal{E}$), while in our proposed algorithm, the largest matrices are of size $2D$. This fact compromises the scalability of the method. For example, with $D=11$ and the data used in Figure \ref{fig:comparison}, we verify that the present algorithm is $\sim 1300$ times slower than our proposed solutions. As for generic Hilbert spaces $\mathcal{H}$, $\mathrm{SEP}\subsetneq\mathrm{PPT}$, the solution of the problem Eq.~\eqref{eq:SDP_Geza} will generally underestimate the physical bound. However, for most practical purposes, e.g. the setting used in Figure~\ref{fig:comparison}, the bound obtained here is compatible with the bound given by our methods. Furthermore, the approach explained here is tailored to linear encoding. Its extension to arbitrary maps is not direct as the algorithm is based on the convex-roof construction Eq.~\eqref{eq:convex_roof} which was only proven valid for unitary transformations (see in particular the last remark in Ref.~\cite{Yu2013}). Note that using the recent results of Ref.~\cite{toth2023quantum}, relevant for calculating the so-called quantum Wasserstein distance of two states, one can actually remove the restriction to symmetric states in the optimization of Eq.~\eqref{eq:SDP_Geza}. \\

\noindent\textbf{Fadel et. al.--} In Ref.~\cite{Fadel2020}, the authors report a method to bound the fidelity with respect to a target state $\hat{\sigma}$ from partial information. As the quantum fidelity is lower-bounded by the Hilbert-Schmidt inner product, $\mathcal{F}(\hat{\rho},\hat{\sigma})\geq \mathrm{Tr}(\hat{\rho}\hat{\sigma})$, the problem can be linearized to an SDP:   

\begin{equation} 
\label{eq:seesaw_janek}
    \begin{array}{crl}
   \mathcal{F}(\hat{\sigma}) \geq  &\min_{\hat{\rho} = \hat{\rho}^\dagger}& \mathrm{Tr}(\hat{\rho}\hat{\sigma}) \\
    &\mbox{s.t.}& \mathrm{Tr}(\hat{\mathbf{D}}\hat{\rho}) = \langle\hat{\mathbf{D}} \rangle \\
    & & \hat{\rho}\succeq 0
    \end{array} \ .
\end{equation}
In the $N$-partite scenario, when the data $\hat{\mathbf{D}}$ are symmetrized few-body correlations (or equivalently, low-order moments of collective observables), the problem can be solved efficiently in $N$ \cite{Aloy2021}. Such low-order moments are sufficient to certify spin-squeezed states generated via one-axis twisting (OAT) dynamics with almost unit fidelity. Within this regime, the QFI of the state would coincide with our bound based on the same moments (see e.g. Figure~\ref{fig:OAT_quad} for $t\lesssim 0.15\pi$). Finally, they also suggest to use the said approach to bound the QFI with respect to transformations generated by collective operators, $\hat{G}^{(N)} = \sum_{i=1}^N\hat{G}_i$ from the fidelity bound $|F_Q[\hat{\rho}, \hat{G}^{(N)}]- F_Q[\hat{\sigma}, \hat{G}^{(N)}]|\leq \zeta\sqrt{1-\mathcal{F}(\hat{\rho},\hat{\sigma})}N^2$, where $\zeta = 8$ in general or $\zeta = 6$ if one of the states is pure \cite{Augusiak2016}. 

\subsection{Exact methods}

Here, we describe the approaches leading to a tight bound, i.e., exactly reproducible by a quantum state. \\

\noindent \textbf{Suggestion by Kołodyński.--}Another possibility is to use the definition of QFI from SLD, $\hat{S}$.  
\begin{equation} 
\label{eq:seesaw_janek}
    \begin{array}{crl}
   F_Q(\hat{G}) = &\min_{\hat{S}, \hat{\rho} = \hat{\rho}^\dagger}& \mathrm{Tr}(i[\hat{G},\hat{S}]\hat{\rho}) \\
    &\mbox{s.t.}&  i[\hat{G},\hat{\rho}] =\{\hat{S},\hat{\rho}\}  \\
    & & \mathrm{Tr}(\hat{\mathbf{D}}\hat{\rho}) = \langle\hat{\mathbf{D}} \rangle\\
    & & \hat{\rho}\succeq 0
    \end{array} \ .
\end{equation}
From Eq.~\eqref{eq:seesaw_janek} we realize that the optimization over the state $\hat{\rho}$ for a given $\hat{S}$ is an SDP. Conversely, for a fixed state $\hat{\rho}$, finding $\hat{S}$ is a linear problem. Consequently, one could alternate between these two problems in a see-saw strategy to find the optimal global solution. In our work, we incorporate implicitly both problems as a single SDP. 

For completeness we shall introduce yet another way to compute the QFI which was derived in Ref.~\cite{Macieszczak2013} from the error propagation formula (c.f. Eq.~\eqref{eq_Cramer_Rao} main text). 
\begin{equation}
    \label{eq:QFI_lin}
    F_Q[\hat{\rho},\hat{G}] = \max_{\hat{X}= \hat{X}^\dagger} -\langle X^2\rangle_{\hat{\rho}} + 2\langle i[\hat{G},\hat{X}] \rangle_{\hat{\rho}}
\end{equation}

The RHS of Eq.~\eqref{eq:QFI_lin} is thus a valid lower bound of the QFI given the data $\hat{\mathbf{D}} = \{\hat{X}^2, i[\hat{G},\hat{X}] \}$. Such approach has been exploited to compute, also with a see-saw strategy, the so-called QFI of a channel $\Lambda$, defined as the optimization problem $\max_{\hat{\rho}}F_Q[\Lambda(\hat{\rho}),\hat{G}]$ over all proper density matrices $\hat{\rho}$. \\

\noindent\textbf{Apellaniz et al.--} We compare our approach with the method introduced in Ref. \cite{apellanizetal2017}. The solution of the central problem of the main text, Eq.~\eqref{eq:central_problem}, is there recovered with the Legendre transform formalism as a maximization over lower bounds, $\inf_\mu \mathcal{W}_{\mu}(\mathbf{r}):=\beta(\mathbf{r})$, where
\begin{equation}
\label{eq:apellaniz}
\mathcal{W}_{\mu}(\mathbf{r}) =\mathbf{r}\cdot\langle\mathbf{D}\rangle + \lambda_{\rm min}\left[(\hat{G}-\mu\mathbb{I})^2 - \mathbf{r} \cdot \hat{\mathbf{D}}\right]  \ ,
\end{equation}
with $\lambda_{\rm min}$ denoting the smallest eigenvalue. For any value of $\mathbf{r}$, $\beta(\mathbf{r})$ will lower bound the minimum QFI with respect to $\hat{G}$  over states $\hat{\rho}$ compatible with $\mathrm{Tr}(\hat{\mathbf{D}}\hat{\rho})=\langle \hat{\mathbf{D}} \rangle$, namely, the solution of the central problem Eq.~\eqref{eq:central_problem} (main text). In other words, $\mathbf{r}$ parameterizes a family of valid lower bounds, the stronger of which will be the $\max_{\mathbf{r}}\beta(\mathbf{r})=\mathrm{QFI}_{\hat{G}}$. As verified in the same Ref.~\cite{apellanizetal2017} one can prove that by construction $\beta$ is tightly coinciding with the solution of the problem. 

While the maximization over $\mathbf{r}$ is convex, this is not the case for $\mu$, which turns the problem nonconvex. Now, for each update of $\mathbf{r}$ in the maximization algorithm, one must search the global minimum on $\mu$ over the large spectral range on $\hat{G}$, $\mu\in [\lambda_{\min}(\hat{G}),\lambda_{\max}(\hat{G})]$. In Figure \ref{fig:landscape_mu}, we show for the same example described in the Appendix \ref{sec:app_dtheta} the landscape which is needed to explore to find the minimum QFI.

\begin{figure}[h]
\centering
   \includegraphics[width=0.6\textwidth]{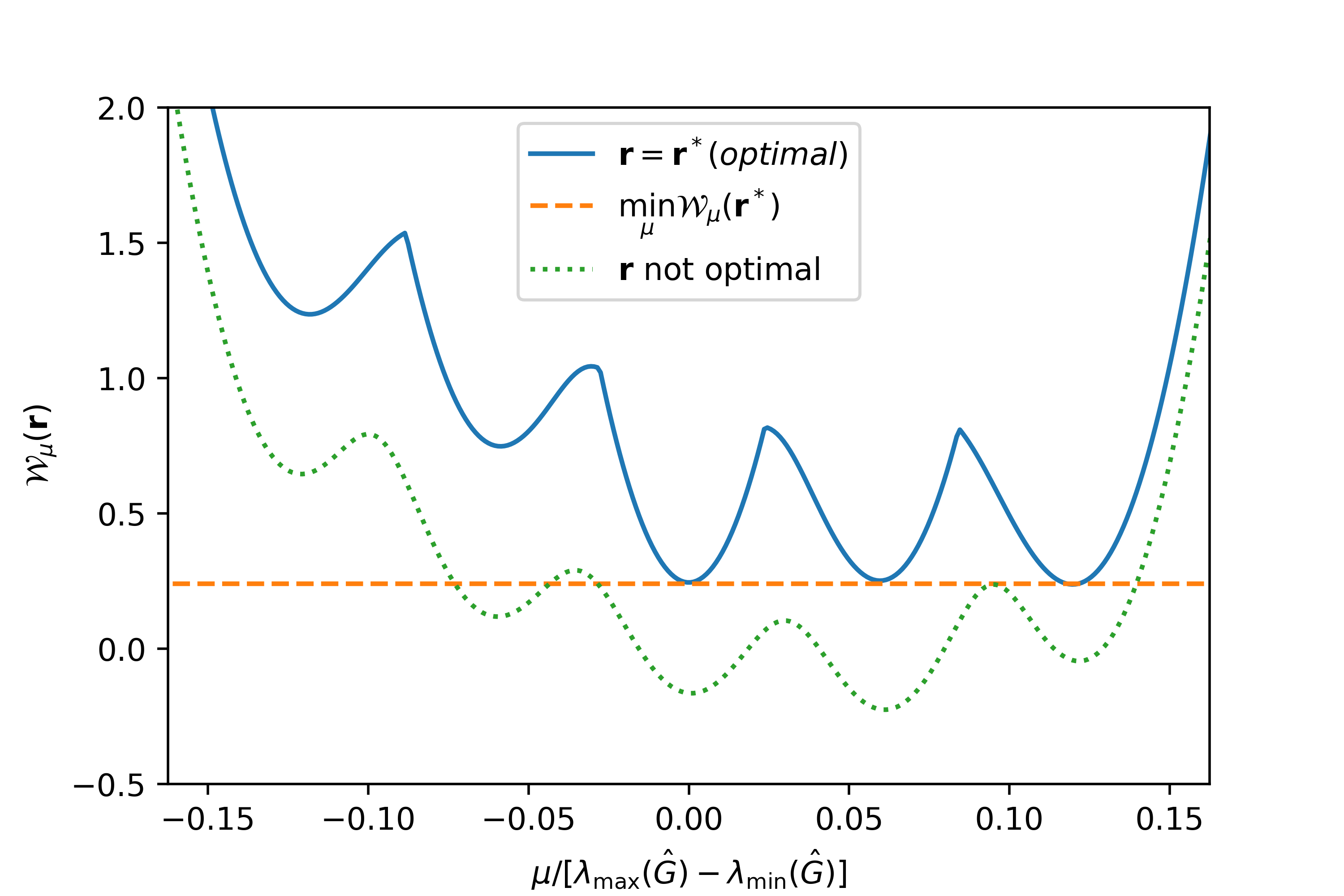} 
	\caption{Blue solid curve $\mathcal{W}_\mu(\mathbf{r})$ of Eq.~\eqref{eq:apellaniz}, for the optimized $\mathbf{r}$ maximizing its global minimum (orange dashed line), $\mathbf{r}^* = \arg\max_{\mathbf{r}}\beta(\mathbf{r})$. For the present example, we have $\mathbf{r}^* = (1.31,-0.95)$. We check that the minimum, $\min_{\mu}\mathcal{W}_{\mu}(\mathbf{r}^*) = 0.2401$, is compatible with our previous value introduced in Appendix \ref{sec:app_dtheta}. In green (dotted), the same for a random (not optimal) setting $\mathbf{r} = (0.89,0.08)$ whose global minimum constitutes generally a lower bound of the QFI.}
	\label{fig:landscape_mu}
\end{figure}

We verify several local minima appearing forcing to sweep over all the range of $\mu$. This fact contrasts with our proposal, which just requires considering an infinitesimal parameter $\delta\theta$. Once more, such method is based on the QFI as a convex roof Eq.~\eqref{eq:convex_roof}, and it is especially suited to evaluate the metrological usefulness in phase estimation tasks such as interferometry. Its generalization to arbitrary parameter-encoding protocols is not addressed in Ref.~\cite{apellanizetal2017}.

\section{Bound for Dicke states of arbitrary spin}
\label{sec:app_obs1}
In this appendix, we prove the observation in the main text regarding the QFI of the manifold of states with $\langle\hat{J}_x^2 \rangle = 0$. First, we notice that such condition implies that the state is an eigenstate of $\hat{J}_x$ with eigenvalue $0$. That is, 
\begin{equation}
\hat{\rho} = \sum_{J}p_J\hat{\Pi}^{(J)}_{J_x = 0} \ ,
\end{equation}
where $\hat{\Pi}^{(J)}_{J_x = 0}$ are projectors properly orthonormalized projecting onto subspaces generated by states with well-defined zero spin projection along $x$, and total spin $J$.  

Next, we use Eq.~\eqref{eq:QFI_formula} to compute the QFI of $\hat{\rho}$ with respect $\hat{J}_z$ with $p = (p_J,0)$ and $\hat{\Pi}^{(p)}= (\hat{\Pi}^{(J)}_{J_x = 0},\hat{\Pi}_{J_x \neq 0})$, where $\hat{\Pi}_{J_x \neq 0} = \mathbb{I}-\sum_{J}\hat{\Pi}^{(J)}_{J_x = 0}$ is the completion of the basis. We can split the double sum as:

\begin{equation}
   \label{eq:FQ_split_Jx0}
    F_Q[\hat{\rho},\hat{J}_z] = 2\sum_{p_J + p_{J'}>0}\frac{(p_J - p_{J'})^2}{p_J + p_{J'}} \mathrm{Tr}(\hat{J}_z\hat{\Pi}^{(J)}_{J_x = 0}\hat{J}_z\hat{\Pi}^{(J')}_{J_x = 0}) + 4\sum_{p_J>0}p_J\mathrm{Tr}(\hat{J}_z\hat{\Pi}^{(J)}_{J_x = 0}\hat{J}_z\hat{\Pi}_{J_x \neq 0})+0 \ .
\end{equation}

We observe that $\hat J_z$ does not connect sectors of different total spin $J$, and therefore $\mathrm{Tr}(\hat{J}_z\hat{\Pi}^{(J)}_{J_x = 0}\hat{J}_z\hat{\Pi}^{(J')}_{J_x = 0}) = 0$ for $J \neq J'$. This implies that the first sum in Eq.~\eqref{eq:FQ_split_Jx0} is zero; and the second one is simply $4\langle\hat{J}_z^2 \rangle = F_{Q}[\hat{\rho},\hat{J}_z]$.

\section{Scaling of the QFI bound with $N$ close to Dicke states}
\label{sec:app_Dicke_scaling}

\begin{figure}[h]
  \centering
    \subfloat[\centering ]{{\includegraphics[width=0.45\textwidth]{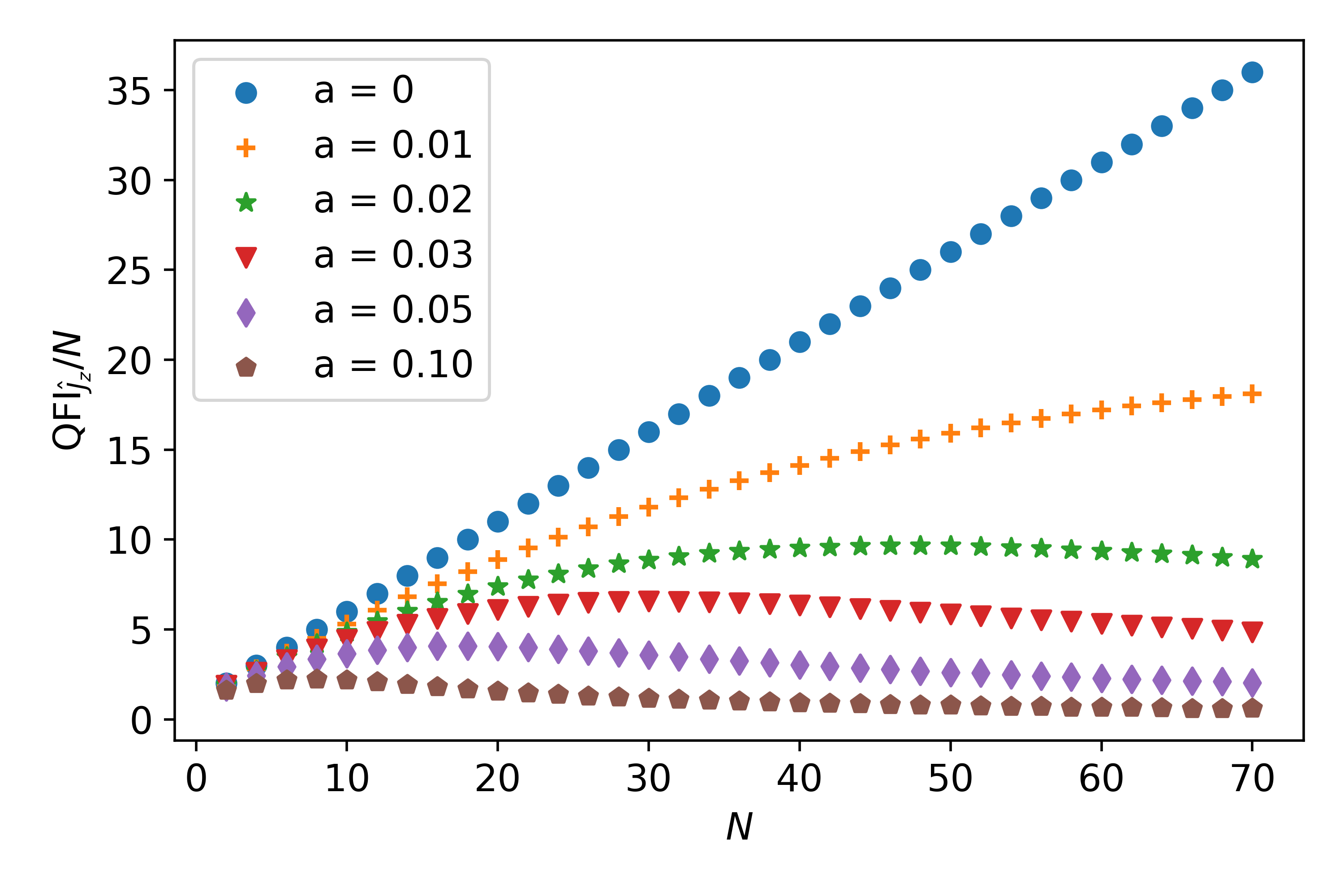} }}%
    \qquad
    \subfloat[\centering ]{{\includegraphics[width=0.45\textwidth]{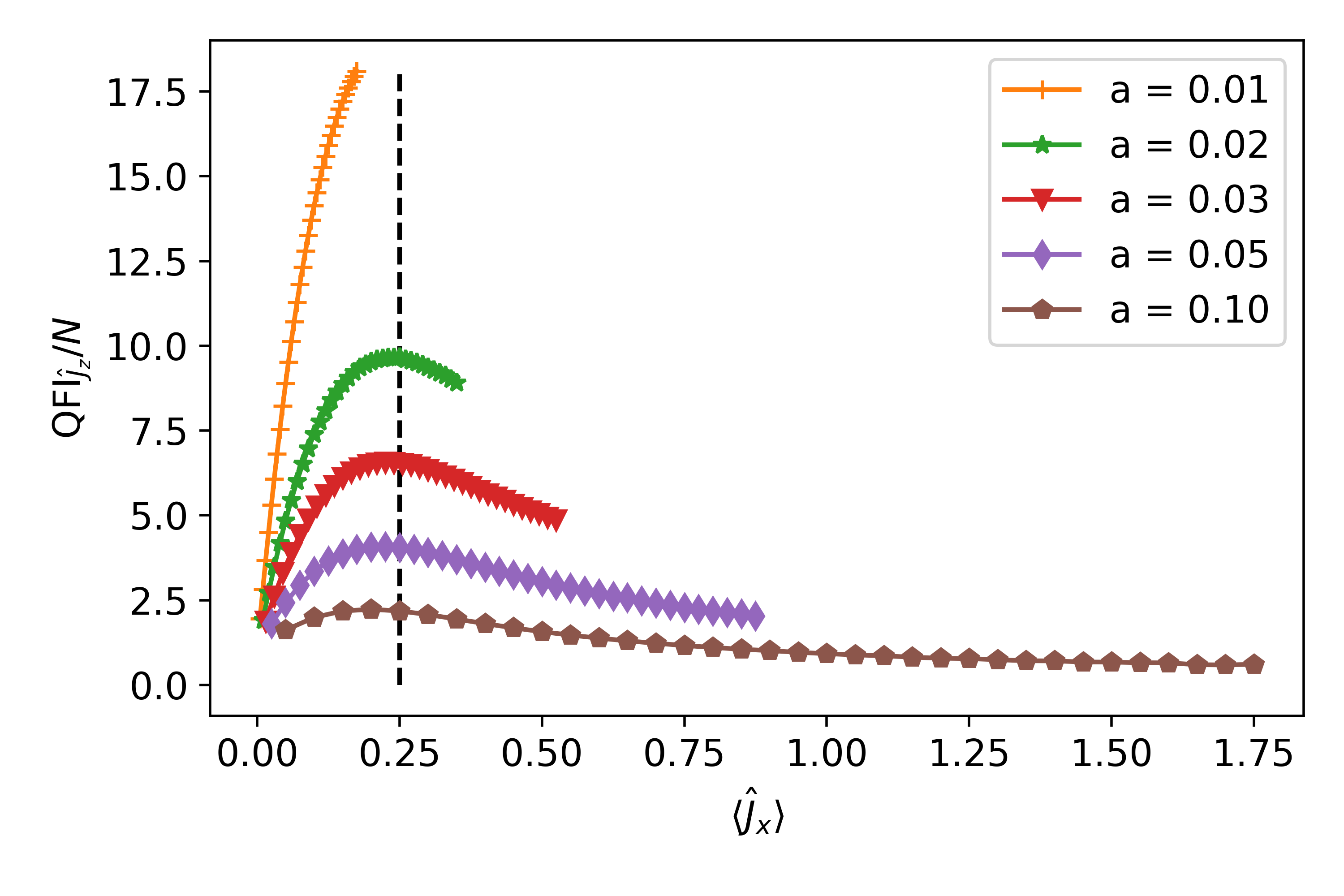} }}%
 \caption{a) Minimum QFI compatible with $\langle\hat{\mathbf{J}}^2\rangle = (N/2)(N/2+1)$ and $\langle\hat{J}_x^2 \rangle = aN/4$ for different values of $a$ and number of particles $N$. In panel b) the same plot for nonzero values of $a$ with the horizontal axis rescaled to $\langle\hat{J}_x^2 \rangle$. The dashed vertical line indicates $\langle\hat{J}_x^2  \rangle = 0.25$.}
 \label{fig:dicke_scaling}
\end{figure}

As detailed in the main text, the condition $\langle \hat{J}_x^2 \rangle = 0$ is quite stringent. Here, we use our approach to find the minimal QFI compatible with some finite $\langle \hat{J}_x^2\rangle$ and $\langle\hat{\mathbf{J}}^2\rangle$. In particular, we fix the total spin to its maximal value, $\langle\hat{\mathbf{J}}^2\rangle = J_{\max}(J_{\max} + 1)$ with $J_{\max}=N/2$, so that we can restrict the optimization to the totally symmetric subspace and reach easily $N = 70$ on a standard computer. This enables us to study the scaling of the QFI with $N$. For $\langle\hat{J}_x^2 \rangle = 0$, we verified the Heisenberg scaling as the QFI is $ \sim\order{J_{\max}^2} = \order{N^2}$. In Figure \ref{fig:dicke_scaling}, we apply a small perturbation $\langle \hat{J}_x^2\rangle = aN/4$ with $a\ll 1$ and study how the QFI bound is modified. 

Figure \ref{fig:dicke_scaling} suggests that the Heisenberg scaling is valid only when $\langle J_x^2\rangle \lesssim 1/4$, where the QFI bound shows a maximum. Beyond this point, the Heisenberg scaling is lost and the metrological usefulness is reduced.

\section{Scaling of the QFI bound with $N$ with respect to squeezing bounds}
\label{sec:app_OAT_scaling}

\begin{figure}[h]
  \centering
    \subfloat[\centering ]{{\includegraphics[width=0.45\textwidth]{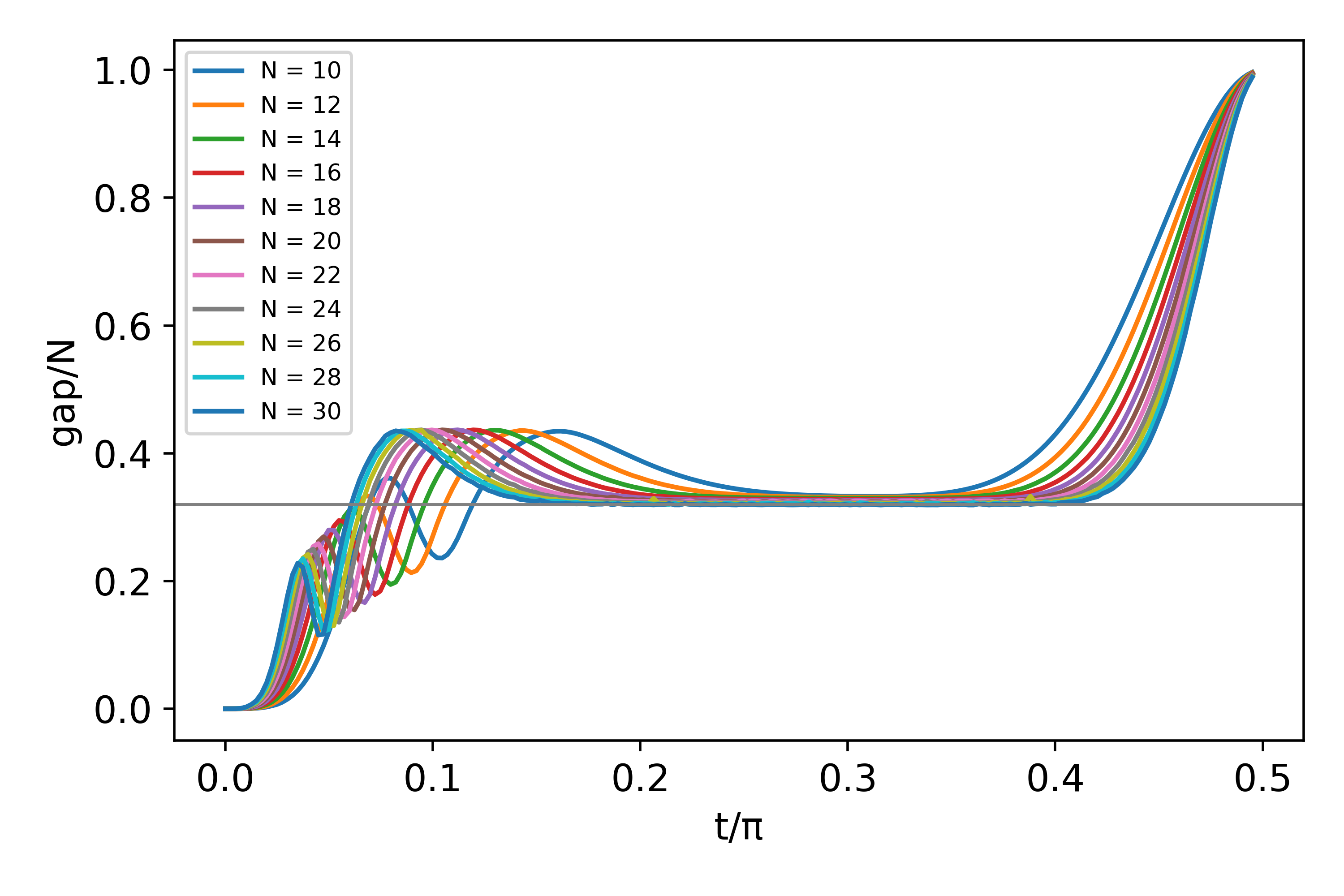} }}%
    \qquad
    \subfloat[\centering ]{{\includegraphics[width=0.45\textwidth]{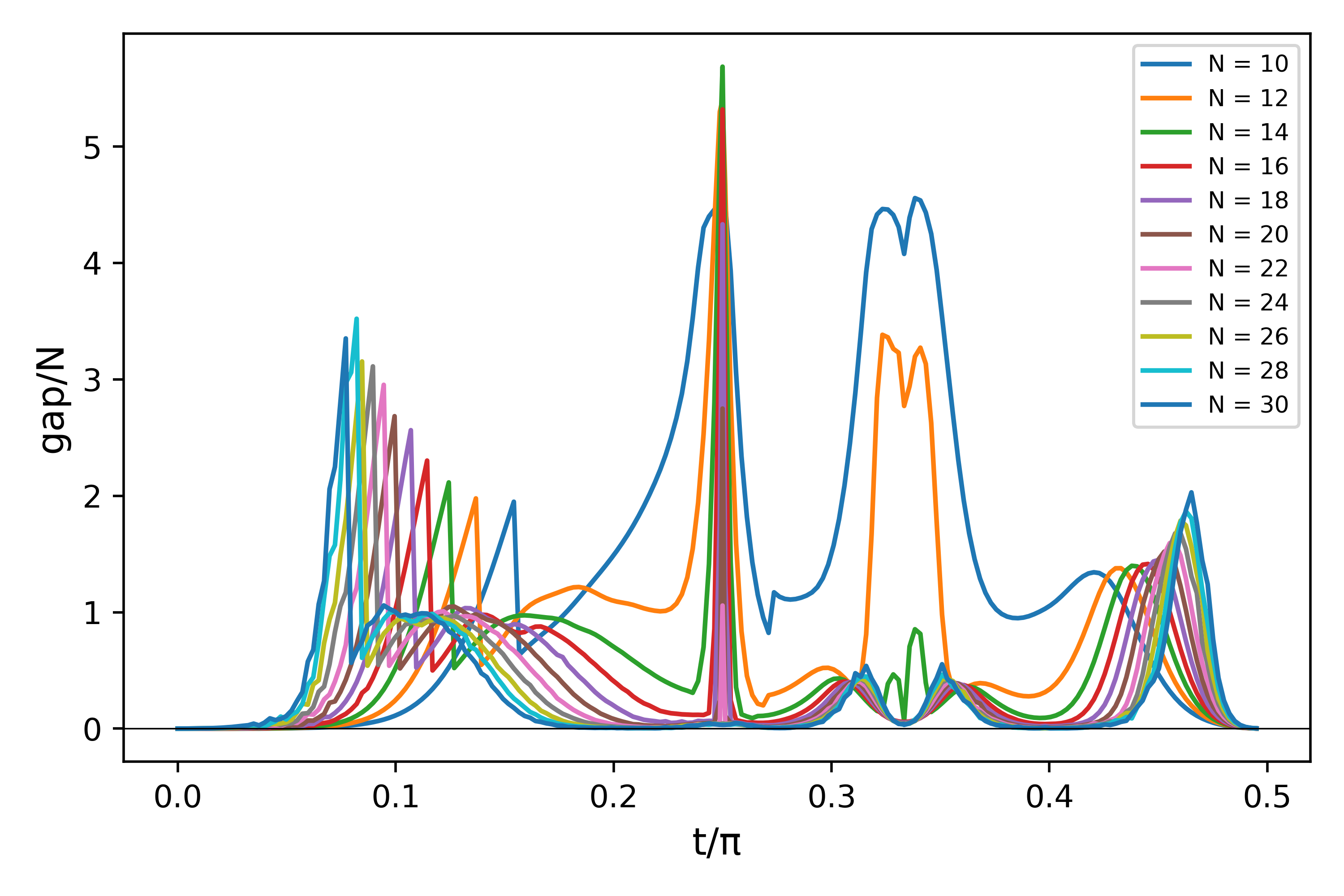} }}%
 \caption{Gap between our SDP-based bound and the analytical squeezing lower bound with the same data is considered: a) up to second moments where we depict also the $\mathrm{QFI} = 0.32N$ horizontal line, b) up to fourth moments, during the OAT dynamics and the line $\mathrm{QFI} = 0$. The squeezing lower bound for second moments corresponds to the conventional spin squeezing parameter. If, in addition, up to fourth moments are taken into account, such bound corresponds to the generalized spin squeezing parameter introduced in Ref. \cite{Gessner2019}.}
 \label{fig:OAT_gap}
\end{figure}

In this appendix, we analyze the scaling of the advantage offered by our method with respect to the analytical lower bound given by the squeezing parameter ensuring the OAT dynamics. In the main text, for $N= 10$ particles, we show a gap existing between such two bounds when up to second moments and fourth moments are considered. Here, we investigate its robustness as the number of parties is increased. Specifically, in Figure \ref{fig:OAT_gap} we display the gap per spin between the two bounds for both cases for $N$ up to 30. 

In Figure \ref{fig:OAT_gap}a, we observe that at intermediate times, the gap is proportional to the particle number, leading to a robust improvement with respect to the usual spin squeezing parameter. At times $t \gtrsim 0.15 \pi$, when the state becomes unpolarized, spin squeezing vanishes while our method certifies metrological usefulness of $\mathrm{QFI}/N \approx 0.32$, which is finite but smaller than the coherent-state limit $\mathrm{QFI}/N = 1$.  

The gap when up to fourth-order moments are considered with respect to the quadratic spin squeezing parameter, introduced in Ref.~\cite{Gessner2019}, is shown in Figure~\ref{fig:OAT_gap}b. Unlike the linear case, we observe at intermediate times the gap saturating to zero as $N$ is increased, signaling no metrological advantage with respect to the squeezing bound, except in the vicinity of the MHCS at $t = \pi/3, \pi/ 4$. It allows us to assess the robustness of the MHCS detection as reported in the main text. On the one hand, the usefulness of the 3-MHCS is already lost for $N= 14$. On the other, the detection of the 4-MHCS is stronger and does not drop below $N$ (coherent limit) until $N=24$. For larger system sizes, higher moments become necessary to detect the metrological usefulness of such states.

\section{Derivation of the data-driven witness $\hat{W}$}
\label{sec:app_witness}

As noticed in the main text, using up to fourth-order moments in the 4-MHCS for $N\leq 10$ (even), the very same state is recovered with unit fidelity as the output of our SDP approach. This suggests that there is only a single state, namely the 4-MHCS, compatible with the data. 

Here, in order to verify this fact, we outline a generic procedure to construct a witness operator $\hat{W}$ which is based on a certain data $\hat{\mathbf{D}}$ (e.g. up to fourth order moments) and whose unique ground state corresponds to a given state $\ket{\Psi}$ (e.g. the 4-MHCS). Without loss of generality, we can set $\lambda_{\min}(\hat{W}) = 0$ as the identity operator is in the data list. Then, we need to find an operator $\hat{W}\in \mathrm{Span}(\hat{\mathbf{D}})$ such that it features $\ket{\Psi}$ as its unique minimal eigenstate. Such problem can be cast as an SDP: 

\begin{equation}
   \label{eq:sdp_W}
   \begin{array}{crl}
     &\max_{\mathbf{n}, \mathrm{cutoff}\geq\Delta \geq 0}& \Delta  \\
    &\mbox{s.t.}& \hat{W} := \mathbf{n}\cdot\hat{\mathbf{D}}  \\
    & & \hat{W}\ket{\Psi} = 0 \\
    & & \ker(\ketbra{\Psi})^\dagger\hat{W} \ker(\ketbra{\Psi}) \succeq \Delta\mathbb{I}
     \end{array} \ ,
\end{equation} 
where $\ker(\ketbra{\Psi})$ projects to the space orthogonal to $\ketbra{\Psi}$. Intuitively, $\Delta$ is the gap between the ground energy $\lambda_{\min}(\hat{W})$ and the first excitation (with multiplicity), which is to be maximized. If the optimal $\Delta$ is zero, it means that the ground state is degenerate, and a witness $\hat{W}$ cannot be found. Conversely, $\Delta>0$ implies that we succeed in finding a witness $\hat{W}$ based on $\hat{\mathbf{D}}$ such that $\ket{\Psi}$ is its unique extremal operator. In this case, we add the constraint $\Delta\leq \mathrm{cutoff}$ to avoid unbounded solutions.

With the above method, we find the existence of such witness for $N\leq 10$ (even). Note that for $N\leq 4$, up to fourth moments data are tomographically complete, so a witness could be simply $\hat{W}=\mathbb{I}-\ketbra{\mathrm{4MHCS}}$. Here, we check that unexpectedly for $N=6,8,10$ fourth moments are also sufficient to determine uniquely the 4-MHCS.

\section{Spin chain}
\label{app_spin_chain}

In this last appendix, we go beyond permutationally-invariant states such as those generated in the OAT dynamics and turn to spatially-structured models. In particular, we consider the nearest-neighbour version of the OAT Hamiltonian studied in the main text, namely the Ising model:
\begin{equation}
   \label{eq:chain}
    \hat{H} = \sum_{l=0}^{N-2}\hat{\sigma}_z^{(l)}\hat{\sigma}_{z}^{(l+1)} \ ,
\end{equation}
where the $N$ spins form a one-dimensional chain with open boundary conditions. Because of the reduced interaction range, the evolution from the initial CSS along $x$ ($\ket{\pi/2,0} = \ket{\rightarrow}^{\otimes N}$) will produce squeezing on a longer timescale than in the OAT model. The spatial structure of the model motivates the use of spatially-structured observables, beyond the collective observables considered in the rest of the paper. Hence, we use structure factor defined as:
\begin{equation}
    \mathcal{C}^{(k)}_{ab} = \sum_{l\neq l'} \langle\hat{\sigma}_a^{(l)} \hat{\sigma}_b^{(l')}\rangle e^{2\pi ik(l-l')/N} \ ,
\end{equation}
where $a,b \in \{x,y,z\}$. The structure factor captures the spin fluctuations at momentum $k\in \{0,1,..,N-1\}$. Indeed, it can be expressed as the second moment of the collective spin at momentum $k$, $\hat{J}^{(k)}_a = (1/2)\sum_{l=0}^{N-1}\hat{\sigma}_a^{(l)}e^{2\pi i kl/N}$, as $\mathcal{C}^{(k)}_{ab}=\langle\hat{J}^{(k)}_a\hat{J}^{(-k)}_b \rangle- N\delta_{ab}/4 $. Such operators generalize to arbitrary momentum $k$ the uniform collective spin used throughout this work, $\hat{\mathbf{J}} = \hat{\mathbf{J}}^{(k=0)}$. 

\begin{figure}[h]
  \centering
   \includegraphics[width=0.5\textwidth]{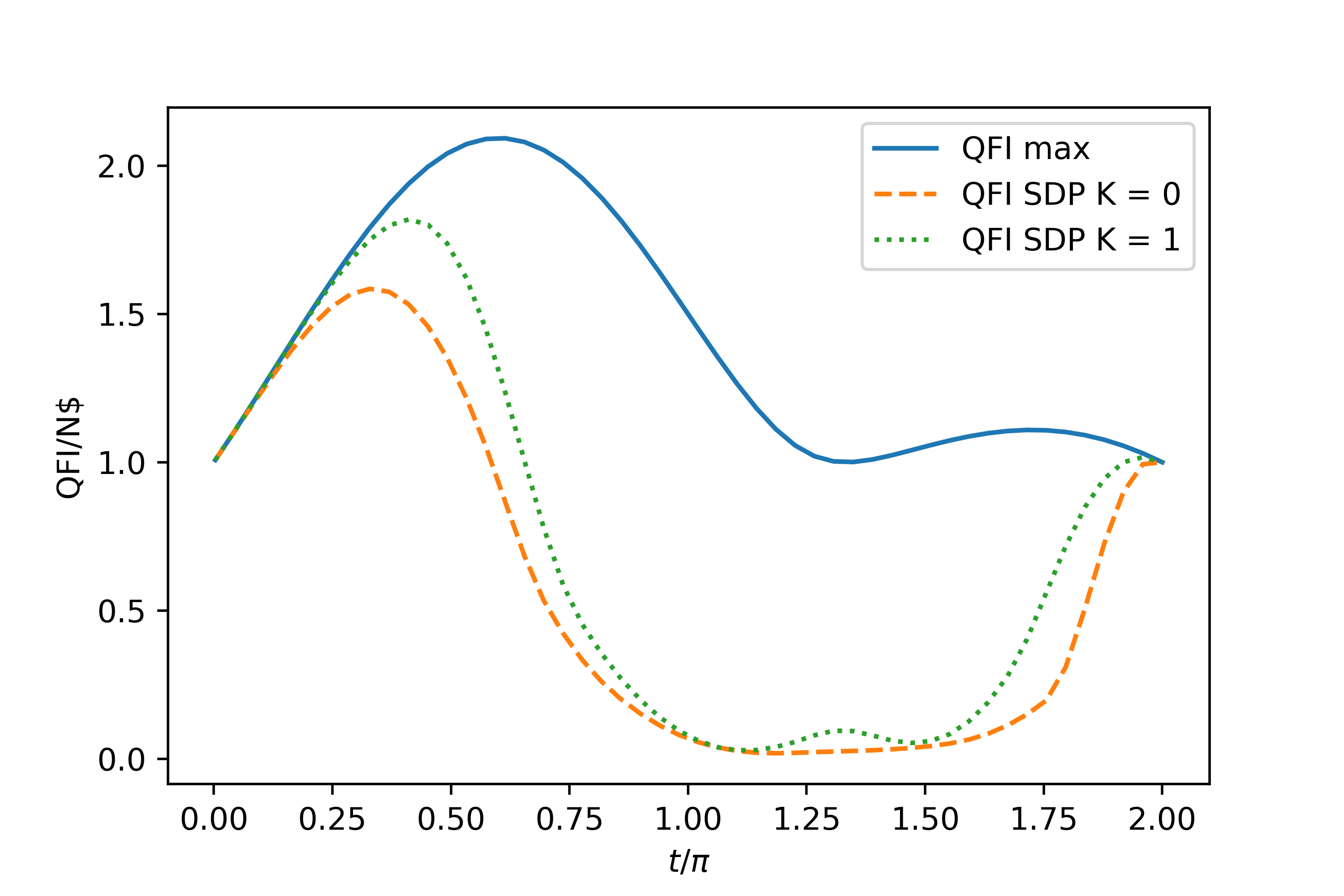} 
	\caption{For $N=4$. Orange (dashed) bound for $K=0$. Green (dotted) bound for $K=1$ (i.e. including $k=0$ and $k=1$). Blue (solid), QFI of the coherent state evolved with $\hat{H}$ Eq.~\eqref{eq:chain}.}
	\label{fig:momentum}
\end{figure}

Here, we will consider data of second moments $\{\mathcal{C}_{ab}^{(k)}\}_{k\leq K}$ only up to momentum $K\leq N-1$ as well as the mean spin $\langle\hat{\mathbf{J}}\rangle$. There is no squeezing parameter based on such data as it is not algebraically closed. Indeed, two momenta $k,k'$ couple to the sum $k + k' \mod N$.  Nevertheless, our algorithm is flexible in the data and we are able to verify that nonzero-momentum observables can improve the metrological bound. In Figure \ref{fig:momentum} we show the results for $N=4$ taking as generator $\hat{J}_{as}$. We observe that the inclusion of momentum $K=1$ indeed improves over the QFI lower bound. This illustrates the potentiality of our approach to certify the metrological usefulness in generic multipartite systems, using spatially-structured data. Note however that in this general case, our algorithm requires computing memory resources scaling exponentially with $N$.

\end{document}